\documentclass[%
 reprint,
 amsmath,amssymb,
 aps,
 pra,
floatfix,
]{revtex4-1}

\usepackage{graphicx}
\usepackage{dcolumn}
\usepackage{bm}
\usepackage{color} 

\begin{document}

\preprint{APS/123-QED}

\title{Excited State Specific Multi-Slater Jastrow Wave Functions}
       
\author{Sergio D. Pineda Flores$^1$}

\author{Eric Neuscamman$^{1,2,}$}%
\email{eneuscamman@berkeley.edu}

\affiliation{
${}^1$Department of Chemistry, University of California, Berkeley, CA, 94720, USA \\
${}^2$Chemical Sciences Division, Lawrence Berkeley National Laboratory, Berkeley, CA, 94720, USA
}

\date{\today}

\begin{abstract}
We combine recent advances in excited state variational principles,
fast multi-Slater Jastrow methods, and selective configuration interaction
to create multi-Slater Jastrow wave function approximations that are
optimized for individual excited states.
In addition to the Jastrow variables and linear expansion coefficients, this 
optimization includes state-specific orbital relaxations in order to avoid the 
compromises necessary in state-averaged approaches.
We demonstrate that, when combined with variance matching to help balance the
quality of the approximation across different states, this approach delivers
accurate excitation energies even when using very modest multi-Slater expansions.
Intriguingly, this accuracy is maintained even when studying a difficult 
chlorine-anion-to-$\pi^*$ charge transfer in which traditional state-averaged 
multi-reference methods must contend with different states that
require drastically different orbital relaxations.
\end{abstract}

\maketitle

\section{Introduction}
\label{sec:intro}
The theoretical study and design of molecular and nano-scale processes driven by photo-absorption remains limited by the accuracy of methods for modeling electronically excited states.
Among the many examples of technological importance in this area, state dependent charge transfer systems like organic solar cells \cite{doi:10.1002/anie.200804709,doi:10.1021/cr900182s} and solar fuel producing catalysts \cite{C2CS35266D,doi:10.1021/nn9015423,doi:10.1021/ar900209b} are of particular note due to their promise for green energy production and their reliance on difficult-to-model charge transfer excited states.
At present, theoretical and computational chemists must choose between methods that are either too expensive to be used in many important charge transfer settings or that have serious shortcomings in their predictive power due to the special challenges that these states pose.
Approaches that can better deal with these challenges, and in particular the strong orbital relaxations that follow a charge transfer excitation, are sorely needed.

While single-excitation theories such as configuration interaction singles (CIS) and time-dependent density functional theory (TDDFT) can be applied in relatively large systems, they preclude the study of double excitations and are often not suitable for dealing with charge transfer.
In the case of CIS, charge transfer suffers from a lack of post-excitation orbital relaxations, \cite{subotnik2011cis_ct} while in TDDFT the challenge lies in balancing local and non-local components of the exchange description. \cite{chaiSystematicLRCDFT}
Linear response methods whose tangent spaces include the doubles excitations, such as equation of motion coupled cluster with singles and doubles (EOM-CCSD) \cite{Krylov:2008:eom_cc_review} are more accurate for charge transfer thanks to the doubles' ability to effect state-specific orbital relaxations for singly excited states.
While one might assume that multi-reference methods such as second order complete active space perturbation theory (CASPT2) would be at least as accurate as single-reference EOM-CCSD for charge transfer, whether they are in practice depends on the effectiveness of the state-averaged (SA) approach to orbital optimization in which one minimizes the average energy of multiple complete active space self consistent field (CASSCF) states.
In cases where the ground and excited states have greatly differing dipoles, as is common for example in charge transfer situations, it is difficult to know a priori that SA orbitals will be equally appropriate for describing the different states.
As we will discuss in one of our examples, the details of how the SA orbitals are arrived at can strongly affect predicted excitation energies in such cases.
The maximum overlap method \cite{gilbert2008mom} for arriving at state-specific orbitals in multi-reference methods can help here, but the stability of this optimization technique is system specific, and so it would be beneficial to develop complementary approaches to address the issue of finding state-specific orbitals for multi-reference excited state wave functions.

Inspired by previous successes \cite{umrigar2007alleviation,giner2013using,scemama2014accurate,giner2015fixed,doi:10.1063/1.1777212,caffarel2016communication,LoosSCIDMC2018,doi:10.1021/acs.jctc.8b00393}
in combining configuration interaction (CI) methods with quantum Monte Carlo (QMC), \cite{foulkes2001quantum}
we investigate a new opportunity for finding excited state specific multi-reference wave functions that has arisen thanks to recent, highly complimentary advances in excited state variational principles,
\cite{Zhao:2016:dir_tar,robinson2017varmatch,zhao2017blocked,Shea:2017:scesvp}
selective configuration interaction methods,
\cite{schriber2016communication,doi:10.1063/1.4955109,doi:10.1021/acs.jctc.6b00407,doi:10.1021/acs.jctc.6b01028,garniron2018selected}
and QMC algorithms for multi-reference wave functions.
\cite{table_method,table_deriv,table_deriv2}
By using variational Monte Carlo (VMC) methods to evaluate the $\hat{H}^2$ term in the objective function of typical excited state variational principles, wave functions can be optimized for specific excited states at a cost that is similar to ground state VMC optimization. \cite{Zhao:2016:dir_tar}
In the case of orbital optimization, this cost has been drastically reduced by recent advances in VMC wave function algorithms, allowing VMC to optimize the orbitals in wave functions containing tens or even hundreds of thousands of Slater determinants. \cite{table_deriv,table_deriv2}
Although determinant expansions of this size are still modest when compared to large CI calculations in quantum chemistry, four factors make this limitation less constraining than it appears.
First, QMC's ability to incorporate weak correlation effects through both Jastrow factors and diffusion Monte Carlo reduces the need for a long tail of small-coefficient determinants in the CI expansion.
\cite{umrigar2007alleviation}
Second, state-specific orbital optimization means that one need not rely on the determinant expansion to correct SA orbitals for the state in question, which is expected to reduce the number of determinants needed to reach a given accuracy.  \cite{doi:10.1063/1.4921984}
Third, the rapid progress in selective CI methodology in recent years provides a highly efficient route to identifying and including only the most important determinants, even in systems and active spaces that are too large to converge with selective CI alone.
Finally, the technique of variance matching \cite{robinson2017varmatch} can help balance the accuracy of different states when faced with these unconverged CI expansions.
Although we will in this study emphasize the ability of this combined methodology to avoid state-averaging and the difficulties it poses in challenging charge transfer situations, the potential applications of multi-reference wave functions with excited state specific orbitals that can capture both weak and strong correlations in systems well beyond the size limits of modern selective CI are clearly very broad and will doubtless merit further exploration.

\section{Theory}
\label{sec:theory}

\subsection{Excited State Specific VMC}

To ensure our orbitals are tailored to the needs of an individual excited state, we will rely on the excited state variational principle that minimizes
\begin{align}
\label{eqn:esvp}
\Omega=\frac{\langle\Psi|(\omega-H)|\Psi\rangle}
            {\langle\Psi|(\omega-H)^2|\Psi\rangle}   
\end{align}
which is a function whose global minimum is the exact Hamiltonian eigenstate with energy immediately above the value $\omega$.  \cite{Zhao:2016:dir_tar}
Just as ground state VMC estimates the energy via a statistical average over $N_s$ position samples $\{\vec{r}_i\}$ drawn from the many-electron probability distribution $p(\vec{r})=|\Psi(\vec{r})|^2/\langle\Psi|\Psi\rangle$,
\begin{align}
&E = \frac{\langle\Psi|H|\Psi\rangle}
            {\langle\Psi|\Psi\rangle} 
  = \int p(\vec{r}) E_L(\vec{r}) d\vec{r} 
  \approx \frac{1}{N_s} \sum_{i=1}^{N_s} E_L(\vec{r}_i), \\
&E_L(\vec{r}) = \frac{H\Psi(\vec{r})}{\Psi(\vec{r})}, \label{eqn:LE}
\end{align}
the objective function $\Omega$ may be statistically estimated as a ratio of two such averages
\begin{align}
    \label{eqn:OmegaStandardGuiding}
    \Omega \approx \frac{\sum_{i=1}^{N_s} \omega-E_L(\vec{r}_i)}
                        {\sum_{i=1}^{N_s} (\omega-E_L(\vec{r}_i))^2}
\end{align}
and minimized via generalizations \cite{Zhao:2016:dir_tar,zhao2017blocked,Shea:2017:scesvp}
of the ground state Linear Method. \cite{UmrTouFilSorHen-PRL-07,doi:10.1063/1.2908237}
Note that, in practice, it is advisable to use a slightly modified probability
distribution from which to draw the samples, a point we will return to in
Section \ref{sec:mgf}.

While this and other \cite{messmer1969variational,messmer1970variational}
excited state variational principles are quite general, one of their most promising
uses is to achieve excited-state-specific relaxations of the orbital basis.
While this goal has been achieved deterministically in both single-determinant wave functions \cite{ye2017sigma}
and linear combinations of single excitations, \cite{shea2018}
as well as by VMC for Jastrow-modified single excitations,
\cite{blunt2017charge,blunt2018charge}
the prospect of extending it to the class of highly-sophisticated
multi-Slater Jastrow (MSJ) wave functions that are directly compatible with VMC and DMC
would greatly increase the potential accuracy that could be sought.
In order to make this goal a reality in a numerically efficient manner, however, we will
need to rely on relatively new techniques for handling MSJ wave functions,
which we will overview in the next two sections.



\subsection{The Table Method}
\label{sec:tableMethod}

First introduced by Clark and coworkers \cite{table_method,morales2012msj}
and recently improved by Filippi and coworkers, \cite{table_deriv,table_deriv2}
the table method has dramatically increased the size of CI expansions that
can be handled by VMC within a MSJ wave function.
While the reader is encouraged to consult the above publications for a fully
detailed explanation of the table method, we will
review the theory here as it will prepare us for the discussion of
orbital optimization, where our approach differs in its details
from previous approaches.
To understand the table method's efficacy, we will analyze the MSJ wave function 
\begin{align}
    \Psi(\vec{r}) = & \hspace{1mm} \psi_{MS}(\vec{r}) \hspace{1mm} \psi_{J}(\vec{r}) \label{eq:1} \\
    \psi_{MS}(\vec{r}) = & \sum_{I=0}^{N} c_{I} D_{I}(\vec{r}) \label{eq:2}
\end{align}
in which $\psi_{J}$ is the symmetric Jastrow correlation factor
and $\psi_{MS}$ the linear combination of antisymmetric Slater Determinants $D_I$.
While in practice one can (and our software does) exploit the factorization
$D_I=D_{I\uparrow}D_{I\downarrow}$ in situations in which the number of
electrons of each spin is fixed, we will for ease of presentation describe
the theory in terms of a fictitious system in which all the electrons are spin
up, in which case Eq.\ (\ref{eq:2}) applies without further factorization of $D_I$.
The generalization to cases where electrons of both spins are present is 
straightforward if a bit tedious.

In our system of $n$ up-spin electrons and $m$ orbitals,
we may define an $n\times m$ matrix $\bm{A}$ whose elements are the
orbital values for each electron's position,
\begin{align}
    A_{i,j} = \phi_j(r_i)
\end{align}
with the first $n$ columns corresponding to the orbitals in the
reference determinant $D_0$.
Note that we are using notation where $\vec{r}$ is a length $3n$ vector of all the
electron coordinates, while $r_i$ is a length $3$ vector of the $i$th electron's coordinates.
We write the determinants in our multi-Slater (MS) expansion as
\begin{align}
    D_I = \mathrm{det}(\bm{A}_I)
\end{align}
where the $n\times n$ matrix $\bm{A}_I$ is formed by taking only those columns of
$\bm{A}$ that correspond to orbitals that are occupied in the $I$th electron configuration.
We will designate the $I=0$ configuration as the reference configuration, typically
the Aufbau configuration, so that, starting from the matrix
$\bm{A}_0$, we may construct the matrix $\bm{A}_I$ via $k_I$ column replacements,
where $k_I$ is the number of single-electron excitations required to
transform configuration $0$ into configuration $I$.
Using two  $n\times k_I$ matrices $\bm{U}_I$ and $\bm{P}_I$, one can express this
relationship as 
\begin{align}
    \bm{A}_I = \bm{A}_0 + \bm{U}_I \bm{P}_I^T.
    \label{eqn:AIfromA0}
\end{align}
Specifically, each column of $\bm{P}_I$ has one element with value $1$ and the
rest zero, while each column of $\bm{U}_I$ contains the difference between the
column of $\bm{A}$ needed for the $I$th configuration and the one it replaces from
the reference configuration.
Noting that $\bm{P}_I^T \bm{P}_I$ is the $k_I\times k_I$ identity matrix,
we can rearrange Eq.\ (\ref{eqn:AIfromA0}) as
\begin{align}
    \bm{U}_I = ( \bm{A}_I - \bm{A}_0 ) \bm{P}_I,
\end{align}
which, along with the matrix determinant lemma, allows us to 
write $D_I$ in terms of $D_0$ and
the determinant of a $k_I\times k_I$ matrix $\bm{\alpha}_I$,
\begin{align}
\vphantom{\Big[ \Big]}
\bm{\alpha}_I &= \bm{P}_I^{T}\bm{A}_{0}^{-1} \bm{A}_I \bm{P}_I \\
D_I &= \mathrm{det}(\bm{A}_{I}) \notag \\
    &= \mathrm{det}(\bm{A}_{0}) \hspace{1mm}
       \mathrm{det}(\bm{I}+\bm{P}_I^{T}\bm{A}_{0}^{-1} \bm{U}_I)
       \notag \\
    &= D_0 \hspace{1mm}
       \mathrm{det}(\bm{I}+\bm{P}_I^{T}\bm{A}_{0}^{-1}
                     ( \bm{A}_I - \bm{A}_0 ) \bm{P}_I)
       \notag \\
    &= D_0 \hspace{1mm} \mathrm{det}(\bm{\alpha}_I). \label{eqn:DI} 
\end{align}
As originally recognized by Clark et al, \cite{table_method}
$\bm{\alpha}_I$ can be constructed efficiently by simply copying the appropriate
elements from the precomputed $n\times m$ ``table'' matrix
\begin{align}
    \bm{T} &= \bm{A}_{0}^{-1} \hspace{0mm} \bm{A}
\end{align}
from whence the table method takes its name.
Thus the cost of evaluating the contribution each additional configuration makes
to the wave function value goes as only $(k_I)^3$, which since in practice $k_I$
tends to be small represents a large speedup compared to the $n^3$
per-configuration cost that would be incurred if the different determinants
$D_I$ were evaluated directly as $\mathrm{det}(\bm{A}_I)$.


Following the presentation of Filippi and coworkers, \cite{table_deriv,table_deriv2}
we can see how this efficiency can be extended to evaluating the local energy
by defining a Jastrow-dependent one-body operator $\hat{O}$.
\begin{align}
    \hat{O}_{i} &= - \frac{1}{2} \left(\frac{\nabla^{2}_{i}\psi_{J}}{\psi_{J}} + \frac{2\nabla_{i}\psi_{J} \cdot \nabla_{i}}{\psi_{J}} + \nabla^{2}_{i} \right)
    \label{eqn:Oi} \\
    \hat{O} &= \sum_{i=1}^{n} \hat{O}_{i}
    \label{eqn:fullO}
\end{align}
By forming the intermediates
\begin{align}
t_I
    &= \sum_{i=1}^{n} - \frac{1}{2}\frac{\nabla^{2}_{i}(\psi_{J}D_{I})}{\psi_{J}D_{I}}  \notag \\
    &= \frac{1}{D_{I}} \sum_{i=1}^{n} - \frac{1}{2}
       \left[   \frac{\nabla^{2}_{i}\psi_{J}}{\psi_{J}} 
              + \frac{2\nabla_{i}\psi_{J} \cdot \nabla_{i}}{\psi_{J}}
              + \nabla^{2}_{i}
       \right] D_{I} \notag \\
     &=  \frac{1}{D_{I}} \sum_{i=1}^{n} \hat{O}_i D_{I}
     \notag \\
     &=  \frac{\hat{O} D_{I}}{D_{I}}  \label{eqn:kineticIntermed}
\end{align}
the kinetic part of the local energy $E_L$ from Eq.\ (\ref{eqn:LE}) can be written as
\begin{align}
    \label{eqn:fastEL}
    K_L = \frac{\sum_I c_I D_I t_I}{\sum_I c_I D_I}.
\end{align}
The kinetic energy intermediates $t_I$ can be converted into
a particularly convenient form by defining the $n\times m$
matrix $\bm{B}$ with elements
\begin{align}
    B_{i,j} = \hat{O}_i \hspace{1mm} \phi_j(r_i),
\end{align}
from which $n\times n$ matrices $\bm{B}_I$ for each configuration
can be constructed in the same fashion as the matrices $\bm{A}_I$
were derived from $\bm{A}$.
Crucially, one can now use the Leibniz formula to rewrite the
intermediates as
\begin{align}
    t_I &= \frac{\hat{O} D_{I}}{D_{I}}  
         = \frac{\partial}{\partial \lambda}
           \ln \big(
             \mathrm{det}(\bm{A}_{I} + \lambda \bm{B}_{I})
           \big)
           \Big|_{\lambda = 0}. \label{eq:111}
\end{align}
Now, for a generic invertable matrix $\bm{G}$ with cofactor matrix
$\bm{C}$, one can use the cofactor formulas for the determinant and
the matrix inverse to arrive at the identity
\begin{align}
	\frac{\partial}{\partial \xi} \ln(\mathrm{det}(\bm{G}))
	&= \sum_{i,j} \frac{\partial \ln(\mathrm{det}(\bm{G}))}{\partial G_{i,j}}
	              \frac{\partial G_{i,j}}{\partial \xi} \notag \\
    &= \sum_{i,j} \frac{C_{i,j}}{\mathrm{det}(\bm{G})}
                  \frac{\partial G_{i,j}}{\partial \xi}  \notag \\
    &= \sum_{i,j} G^{-1}_{j,i} 	\frac{\partial G_{i,j}}{\partial \xi} \notag \\
    &= \mathrm{Tr}[\bm{G}^{-1}\frac{\partial \bm{G}}{\partial \xi}]. \label{eq:04}
\end{align}
For the reference configuration, this identity gives us
\begin{align}
    t_0 &= \mathrm{Tr}[\bm{A}_{0}^{-1}\bm{B}_{0}]. \label{eqn:t0Tr}  
\end{align}
For the other configurations, we note that Eq.\ (\ref{eqn:DI}) remains
valid under the replacement $\bm{A} \rightarrow  \bm{A} + \lambda \bm{B}$,
which we use with Eqs.\ (\ref{eq:111}) and (\ref{eq:04})
to find that
\begin{align}
    t_I &= \frac{\partial}{\partial \lambda}
           \Big[ \ln \big(
             \mathrm{det}(\bm{A}_{0} + \lambda \bm{B}_{0})
           \big) \notag \\
          & \hspace{3mm} + \ln \Big( \mathrm{det}\big(
          \bm{P}^T_I
          (\bm{A}_0 + \lambda \bm{B}_0)^{-1}
          (\bm{A}_I + \lambda \bm{B}_I)
          \bm{P}_I
          \big) \Big)
           \Big]_{\lambda = 0} \notag \\
        &= t_0 + \mathrm{Tr}[ \bm{\alpha}_I^{-1} \bm{\beta}_I ]
        \label{eqn:tITr}
\end{align}
where we have defined the $k_I\times k_I$ matrix
\begin{align}
    \bm{\beta}_I &= \frac{\partial}{\partial \lambda}
                    \Big(
                    \bm{P}^T_I
                    (\bm{A}_0 + \lambda \bm{B}_0)^{-1}
                    (\bm{A}_I + \lambda \bm{B}_I)
                    \bm{P}_I
                    \Big) \Big|_{\lambda=0}
    \notag \\ 
                &=  \bm{P}^T_I (  \bm{A}_0^{-1} \bm{B}_I
                  - \bm{A}_0^{-1} \bm{B}_0 \bm{A}_0^{-1} \bm{A}_I ) \bm{P}_I.
\end{align}
As for $\bm{\alpha}_I$ and $\bm{T}$, we can define a second table matrix
\begin{align}
    \bm{Z} = \bm{A}_0^{-1} \bm{B} - \bm{A}_0^{-1} \bm{B}_0 \bm{A}_0^{-1} \bm{A}
\end{align}
such that each $\bm{\beta}_I$ can be built by simply copying the appropriate
elements from the precomputed matrix $\bm{Z}$.
Combining Eqs.\ (\ref{eqn:DI}), (\ref{eqn:fastEL}), and (\ref{eqn:tITr}) leads
us to our final expression for the kinetic portion of the local energy
\begin{align}
  \label{eqn:finalKL}
  K_L = t_0
        + \frac{\sum_{I=1}^N c_I \hspace{0.5mm} \mathrm{det}(\bm{\alpha}_I)
                    \hspace{0.5mm} \mathrm{Tr}[\bm{\alpha}_I^{-1}\bm{\beta}_I]}
        {c_0 + \sum_{I=1}^N c_I \hspace{0.5mm} \mathrm{det}(\bm{\alpha}_I)}
\end{align}
in which the local kinetic energy $t_0$ of the reference-configuration-based
single-Slater-Jastrow wave function is corrected by the second term to
produce the local kinetic energy of the full MSJ wave function.
We therefore see that the table method allows the local energy 
to be evaluated for a cost that goes as $n^2 m$ for the construction of
$\bm{T}$, $\bm{Z}$, and $t_0$ plus an additional
per-configuration cost that goes as just $(k_I)^3$.

\subsection{MSJ Orbital Optimization}
\label{sec:msjoo}

To achieve state-specific orbital optimization, we minimize the excited
state variational principle from Eq.\ (\ref{eqn:esvp}) via a generalization
of the linear method. \cite{Shea:2017:scesvp}
In practice, this requires evaluating the derivatives of $E_L$ and $\ln(\Psi)$
with respect to the wave function variables
at every sample of the electron positions, an endeavor that has recently
been made drastically more efficient thanks to the approach of 
Filippi and coworkers. \cite{table_deriv,table_deriv2}
Although the details of our approach to these derivatives differ from theirs, 
the basic idea of creating efficient intermediates is shared and
the resulting cost scaling is the same.

To begin, we recognize that, for a given set of electron positions, both
$E_L$ and $\ln(\Psi)$ are many-input/single-output functions of the
wave function's variational parameters, and so we expect that the automatic
differentiation approach of reverse accumulation
\cite{shea2018,Griewank-Walther-book,sorella2010qmcforces,
      Neuscamman2013jagp,Aspuru-Guzik:2017:ad}
will yield all the
necessary derivatives for a cost that is a small constant multiple
of the cost of evaluating $E_L$ and $\ln(\Psi)$.
Reverse accumulation is essentially a careful exploitation of the chain
rule, and so, using Eq.\ (\ref{eqn:finalKL}) and
remembering that $\bm{\alpha}_I$ and $\bm{\beta}_I$ are
$k_I\times k_I$ matrices with elements copied, respectively, from $\bm{T}$ and $\bm{Z}$,
we formulate our local energy derivatives as
\begin{align}
    \frac{\partial E_L}{\partial \mu} =
    \frac{\partial t_0}{\partial \mu}
 &+ \mathrm{Tr}\left[\left(\frac{\partial K_L}{\partial \bm{T}}\right)^T
                       \frac{\partial \bm{T}}{\partial \mu} \right]
    \notag \\
 & \qquad \qquad
 + \mathrm{Tr}\left[\left(\frac{\partial K_L}{\partial \bm{Z}}\right)^T
                       \frac{\partial \bm{Z}}{\partial \mu} \right]
 \label{eqn:LEderivs}
\end{align}
in which $\mu$ is an as-yet unspecified orbital rotation variable.
An inspection of Eq.\ (\ref{eqn:finalKL}) reveals that, while every configuration $I$
makes a contribution to the matrix $\partial K_L/\partial\bm{Z}$,
each of these individual contributions affects only $(k_I)^2$ of its matrix elements
because $\bm{\beta}_I$ was built by copying only that many elements from $\bm{Z}$.
Using 
\begin{align}
    \frac{\partial}{\partial \bm{\beta}_I} \mathrm{Tr}[\bm{\alpha}_I^{-1} \bm{\beta}_I]
    = (\bm{\alpha}_I^{-1})^T
\end{align}
and the fact that the values of $\mathrm{det}(\bm{\alpha}_I)$ and
$\bm{\alpha}_I^{-1}$ are known already from our evaluation of $K_L$,
we find that the additional per-configuration cost
(beyond that required to evaluate $K_L$ itself)
of constructing $\partial K_L/\partial\bm{Z}$ goes as just $(k_I)^2$.
By a similar chain of logic, the relationships
\begin{align}
    \frac{\partial}{\partial \bm{\alpha}_I} \mathrm{Tr}[\bm{\alpha}_I^{-1} \bm{\beta}_I]
    &= -(\bm{\alpha}_I^{-1} \bm{\beta}_I \bm{\alpha}_I^{-1})^T  \\
    \label{eqn:detAlphaDeriv}
    \frac{\partial}{\partial \bm{\alpha}_I} \mathrm{det}(\bm{\alpha}_I)
    &= \mathrm{det}(\bm{\alpha}_I) (\bm{\alpha}_I^{-1})^T
\end{align}
lead to the conclusion that the additional per-configuration cost of constructing $\partial K_L/\partial\bm{T}$ goes as $(k_I)^3$
due to the need to form the $\bm{\alpha}_I^{-1} \bm{\beta}_I \bm{\alpha}_I^{-1}$
matrix products.
We therefore see that the reverse-accumulation intermediate matrices
$\partial K_L/\partial\bm{T}$ and $\partial K_L/\partial\bm{Z}$
can be evaluated for an additional cost that has the same scaling
as $K_L$ itself and, in practice, a smaller prefactor due to the fact
that matrix-matrix multiplication is typically faster than matrix inversion.

Once $\partial K_L/\partial\bm{T}$ and $\partial K_L/\partial\bm{Z}$ have been
constructed, the remaining cost of evaluating the orbital rotation derivatives
is independent of the number of configurations in the CI expansion.
The term $\partial t_0 / \partial \mu$ is simply the orbital-rotation
derivative for a single-Slater-Jastrow wave function,
and the cost to evaluate this term for all the orbital rotation variables is known
to be cubic with system size \cite{table_deriv}
assuming that both the number of orbitals $m$
and electrons $n$ grow linearly with system size.
This leaves the question of the cost to construct
$\partial \bm{T}/\partial\mu$ and $\partial \bm{Z}/\partial\mu$,
for which we will need to be explicit as to the parameterization of
our orbital rotations.
A general rotation among the molecular orbitals can be achieved via a unitary
transformation
\begin{align}
    \bm{A} \rightarrow \bm{A} \bm{U} \qquad \qquad \bm{B} \rightarrow \bm{B} \bm{U}
\end{align}
in which the unitary $m\times m$ matrix $\bm{U}$ is parameterized by
the exponential of an antisymmetric matrix $\bm{X}=-\bm{X}^T$,
\begin{align}
    \bm{U} = \exp (\bm{X}).
\end{align}
If we define $\bm{X}=0$ for the current molecular orbitals, then the derivatives
we need for the $\bm{T}$ matrix can be derived as
\begin{align}
    \frac{\partial T_{a,b}}{\partial X_{i,j}} \Big|_{\bm{X}=0}
    = \delta_{j,b} \hspace{0.5mm} T_{a,i} 
       - \Theta\left(n+\frac{1}{2}-j\right) \delta_{a,i} \hspace{0.5mm} T_{j,b}
\end{align}
where the Heaviside function $\Theta$ ensures that the second term only contributes
when $j\in \{1,2,\ldots,n\}$.
This results in an efficient formulation for the middle term from
Eq.\ (\ref{eqn:LEderivs})
\begin{align}
 \label{eqn:TDerivsForKL}
 & \mathrm{Tr}\left[\left(\frac{\partial K_L}{\partial \bm{T}}\right)^T
                       \frac{\partial \bm{T}}{\partial X_{i,j}} \right]
 \notag \\[2mm]
 & \hspace{6mm} = \begin{cases}
     \left[ \bm{T}^T \frac{\partial K_L}{\partial \bm{T}}
      - \frac{\partial K_L}{\partial \bm{T}} \bm{T}^T \right]_{i,j}
    &  \mathrm{if} \hspace{2mm} i,j \in \{1,2,\ldots,n\} \\[4mm]
     \left[ \bm{T}^T \frac{\partial K_L}{\partial \bm{T}}
     \right]_{i,j}
    &  \mathrm{otherwise}
   \end{cases}
\end{align}
which we see can be evaluated for a cost that grows cubically with
the system size.
Although we will omit the details for brevity, the $\bm{Z}$ term
(i.e.\ the third term) in Eq.\ (\ref{eqn:LEderivs}) can also
be formulated in a cubic-cost way once we have the
intermediate matrix $\partial K_L/\partial\bm{Z}$,
and so we find that all three parts of the local energy derivatives
with respect to the orbital rotation variables $\bm{X}$ can
be evaluated at a cost whose scaling is the same as $E_L$ itself.

For the wave function derivatives $\partial \ln \Psi / \partial X_{i,j}$ we start
by combining Eqs.\ (\ref{eq:2}) and (\ref{eqn:DI}) to give
\begin{align}
    \label{eqn:PsiForDeriv}
    \ln(\Psi_{MS}) =
    \ln(D_0)+ \ln\left( c_0 + \sum_{I=1}^N c_I \hspace{1mm} \mathrm{det}(\bm{\alpha}_I) \right)
\end{align}
which, applying the chain rule as before, gives the derivatives with respect
to orbital rotation variables as
\begin{align}
    \frac{\partial \ln(\Psi)}{\partial X_{i,j}} =
    \frac{\partial \ln(D_0)}{\partial X_{i,j}}
 &+ \mathrm{Tr}\left[\left(\frac{\partial \ln(\Psi)}{\partial \bm{T}}\right)^T
                       \frac{\partial \bm{T}}{\partial X_{i,j}} \right].
 \label{eqn:lnPsiDeriv}
\end{align}
We again recognize that the first term here is the same as for the single-Slater-Jastrow
case and that its contribution to all the derivatives
$\partial\ln(\Psi)/\partial X_{i,j}$ can be evaluated for a cost that grows
cubically with system size. \cite{table_deriv}
For the second term, as for the derivatives $\partial K_L/\partial\bm{T}$ above,
each configuration $I$
contributes to only $(k_I)^2$ elements of $\partial \ln(\Psi)/\partial\bm{T}$,
and since the inverse matrices $\bm{\alpha}_I^{-1}$ needed by
Eq.\ (\ref{eqn:detAlphaDeriv})
have already been evaluated in the course of evaluating $E_L$,
the per-configuration cost of constructing $\partial \ln(\Psi)/\partial\bm{T}$
will go as just $(k_I)^2$.
Once $\partial \ln(\Psi)/\partial\bm{T}$ is built, its contribution to
Eq.\ (\ref{eqn:lnPsiDeriv}) will be the same as Eq.\ (\ref{eqn:TDerivsForKL}),
but with $\partial K_L/\partial\bm{T}$ replaced by $\partial \ln(\Psi)/\partial\bm{T}$.
We therefore see that, like the local energy, the logarithmic wave function
derivatives can be evaluated at an additional cost that has the same scaling
as the table method's fast evaluation of the wave function itself.

If we are given a set of $N_s$ samples of the electron positions,
we can therefore evaluate the wave function value, the local energy,
and their derivatives with respect to all orbital rotation variables
for a per-sample cost with two parts:
a per-configuration cost that goes as $(k_I)^3$ and a
configuration-independent cost that grows cubically with system size.
However, in practice, the set of samples must be generated somehow,
and this step is typically accomplished by using the Metropolis
algorithm to propagate a Markov chain based on one-electron moves.
To mitigate auto-correlation, it is usually necessary to separate
samples at which the local energy is evaluated with a number of
one electron moves that grows linearly with the size of the
system (e.g.\ one might make one move per electron in between local
energy evaluations).
This Markov chain propagation turns out to have a worse scaling than
the evaluation of the local energies themselves, at least in the
absence of pseudopotentials.
While the matrices that the evaluations of $\Psi$ and $K_L$ depend on
--- namely $\bm{A}$, $\bm{B}$, $\bm{T}$, and $\bm{Z}$ ---
can all be updated using the Sherman Morrison formula in a way that
lowers their per-move cost to be only quadratic in the system size,
there is still the loop over the $N$ configurations in the CI expansion
to consider.
This loop must be performed for each one-electron move, and so,
if we assume that the number of orbitals $m$ is proportional to
the number of electrons $n$ and that the maximum excitation level
is $k_{\mathrm{max}}$, we find that the per-sample cost of propagating
the Markov chain scales as $\mathcal{O}(n^3 + n N k_{\mathrm{max}}^3)$
even though the per-sample cost (without pseudopotentials) of then
evaluating the local energy at each sample goes as only
$\mathcal{O}(n^3 + N k_{\mathrm{max}}^3)$.
In practice, of course, pseudopotentials are quite common, and since
for each sample they require evaluating $\Psi$ at a number of grid
points that grows linearly with the number of atoms,
we find that the per-sample cost of the local energy with
pseudopotentials (assuming the number of atoms is proportional to $n$)
has the same $\mathcal{O}(n^3 + n N k_{\mathrm{max}}^3)$ scaling
as the Markov chain propagation.

\subsection{The Jastrow Factor}
\label{sec:jastrow}

Although a wide variety of forms for the Jastrow factor $\Psi_J$ can be used
efficiently in the table method via Eqs.\ (\ref{eqn:Oi})-(\ref{eqn:kineticIntermed}),
we have for this study kept this component relatively simple by employing
only one- and two-body Jastrows.
Our Jastrow takes the form
\begin{align}
	\psi_J &= \mathrm{exp} \big(\hspace{1mm}
	           J_{1}(\vec{R},\vec{r}) + J_{2}(\vec{r})
	         \hspace{1mm}\big) \label{eq:jastrow} \\[2mm]
  J_{1} &= \sum_{k} \sum_{i} \chi_{k}(|r_{i}-R_{k}|) \label{eqn:J1} \\
  J_{2} &= \sum_{i} \sum_{j>i} u_{ij}(|r_{i}-r_{j}|) \label{eqn:J2}
\end{align}
where $\vec{R}$ and $\vec{r}$ are the vectors giving the nuclear and electron
coordinates, respectively.
The function $u_{ij}$ takes on one of two forms, $u_{ss}$ or $u_{os}$,
depending on whether electrons $i$ and $j$ have the same spin or opposite
spins, and these two forms are constructed using splines \cite{KimQMCPACK2018}
so as to guarantee that the appropriate electron-electron cusp
conditions are satisfied.
The functions $\chi_k$ are similarly constructed of splines \cite{KimQMCPACK2018}
and can either be formulated to enforce the nuclear cusp condition or to be cusp-free
in cases where either a pseudopotential is used or the cusp is build into the orbitals.


\subsection{Modified Guiding Function}
\label{sec:mgf}

Although ground state VMC often draws samples from the probability
distribution $|\Psi|^2/\langle\Psi|\Psi\rangle$ due to the allure of
the zero variance principle, \cite{foulkes2001quantum}
this approach is not statistically robust
when estimating the energy variance,
\begin{align}
    \label{eqn:variance}
    \sigma^2 = \frac{\langle\Psi|(H-E)^2|\Psi\rangle}{\langle\Psi|\Psi\rangle}.
\end{align}
The trouble comes from the fact that the local energy $H\Psi/\Psi$ can diverge,
because $\Psi$ can be zero when $\nabla^2\Psi$ is not.
Although this divergence is integrable for $E$ and $\sigma^2$ and so poses no
formal issues for estimating $E$, it is not integrable
for the variance of $\sigma^2$,
\cite{Trail:2008:heavy_tail,Trail:2008:alt_sampling,robinson2017varmatch}
and so a naive approach in which samples are drawn from
$|\Psi(\vec{r}\hspace{0.5mm})|^2/\langle\Psi|\Psi\rangle$
will not produce normally distributed estimates for $\sigma^2$.
Due to the relationship
\begin{align}
  \Omega(\Psi)
    = \frac{\langle\Psi|(\omega -\hat{H})|\Psi\rangle}
           {\langle\Psi|(\omega -\hat{H})^2|\Psi\rangle}
    = \frac{\omega-E}{(\omega-E)^2+\sigma^2}
    \label{eqn:OmegaAsEandSigma}
\end{align}
we are left with the consequence that statistical estimates for $\Omega$
via Eq.\ (\ref{eqn:OmegaStandardGuiding}) will also not
be normally distributed.

In a previous study, \cite{robinson2017varmatch} we overcame this difficulty with the
alternative importance sampling function
\begin{align}
  \label{eqn:mgf}
  |\Psi_{\mathrm{M}}|^2 = |\Psi|^{2}
    + \frac{\epsilon \hspace{0.5mm} |\nabla^{2}\Psi|^{2}}
           {1 + \mathrm{exp}
              \Big[ 
              \big(\ln|\Psi| - \overline{\ln|\Psi|} + \sigma_{\overline{\Psi}}
              \hspace{0.5mm} \big)
              \hspace{0.5mm} / \hspace{0.5mm}
              \sigma_{\overline{\Psi}} \Big] }
\end{align}
in which the average
(\hspace{0.5mm}$\overline{\ln|\Psi|}$\hspace{0.5mm})
and standard deviation
($\sigma_{\overline{\Psi}}$) of the logarithm of the wave function
absolute value are estimated on a short sample
drawn from the traditional distribution
$|\Psi(\vec{r}\hspace{0.5mm})|^2/\langle\Psi|\Psi\rangle$.
For any $\epsilon>0$, Eq.\ (\ref{eqn:mgf}) guarantees that
the wave function ratio 
\begin{align}
    \label{eqn:wfr}
    \eta(\vec{r}\hspace{0.5mm})
    &= \big|   \Psi(\vec{r}\hspace{0.5mm})
              / \Psi_{\mathrm{M}}(\vec{r}\hspace{0.5mm}) \big|
\end{align}
and the modified local energy
\begin{align}
    \label{eqn:mle}
    E_L^{\mathrm{M}}(\vec{r}\hspace{0.5mm})
    &= \eta(\vec{r}\hspace{0.5mm}) E_L(\vec{r}\hspace{0.5mm})
\end{align}
will be finite everywhere, which implies that if we draw $N_s$
samples from the distribution
$|\Psi_{\mathrm{M}}(\vec{r}\hspace{0.5mm})|^2
 / \langle\Psi_{\mathrm{M}}|\Psi_{\mathrm{M}}\rangle$,
the resulting statistical estimates
\begin{align}
    E &\approx \frac{ \sum_{i=1}^{N_s} \eta(\vec{r}_i) E_L^{\mathrm{M}}(\vec{r}_i)}
                    { \sum_{i=1}^{N_s} \big(\eta(\vec{r}_i)\big)^2 }
    \label{eqn:mgfEnergy} \\[2mm]
    \sigma^2 
     &\approx \frac{ \sum_{i=1}^{N_s}
                       \Big( 
                         E_L^{\mathrm{M}}(\vec{r}_i) - \eta(\vec{r}_i) E
                       \Big)^2
                   }{ \sum_{i=1}^{N_s} \big(\eta(\vec{r}_i)\big)^2 }
    \label{eqn:mgfSigma} \\[2mm]
    \Omega 
     &\approx \frac{ \sum_{i=1}^{N_s}
                       \eta(\vec{r}_i)
                         \Big( \omega \hspace{0.5mm} 
                               \eta(\vec{r}_i) - E_L^{\mathrm{M}}(\vec{r}_i) \Big)
                   }
                   { \sum_{i=1}^{N_s}
                       \Big( 
                         \omega \hspace{0.5mm} \eta(\vec{r}_i)
                         - E_L^{\mathrm{M}}(\vec{r}_i)
                       \Big)^2
                   }
    \label{eqn:mgfOmega}
\end{align}
are guaranteed to be normally distributed for sufficiently large $N_s$.
Note that, due to divergences in $E_L$, this central limit theorem
guarantee would not be true \cite{Trail:2008:heavy_tail,Trail:2008:alt_sampling}
for the $\sigma^2$ and $\Omega$ estimates if we had made the traditional choice
of $|\Psi_{\mathrm{M}}|^2=|\Psi|^2$.
Also note that, although the denominator in Eq.\ (\ref{eqn:mgf})
is not strictly necessary in
order to recover normal statistics for the estimates of $\sigma^2$
and $\Omega$, it does help keep us as close as possible to
$|\Psi|^2$ and thus the zero variance principle by smoothly
switching off the modification when the value of the wave function
magnitude is large relative to its average.
This way, the divergences that occur near the nodes of $\Psi$ are avoided,
while at the same time the probability distribution is left essentially
unmodified in regions of space where the wave function magnitude is large.

By exploiting the table method, it is possible to employ $\Psi_M$ without changing
the overall cost scaling of the Markov chain propagation.
Although we now must evaluate $\nabla^2\Psi = -2\Psi K_L$ every time
we move one of our electrons,
we saw in Sections \ref{sec:tableMethod} and \ref{sec:msjoo}
that evaluating the local kinetic energy $K_L$
has the same cost scaling as that of evaluating $\Psi$ itself
once the matrices $\bm{A}$, $\bm{B}$, $\bm{T}$, and $\bm{Z}$ have
been prepared.
Thanks to the Sherman Morrison formula, these matrices can be updated
efficiently during each one-electron move, and although the
new per-move need for $\bm{Z}$
and $K_L$ does increase the update cost, it does not change the scaling.
Overall, our experience has been that the practical benefits
of using $\Psi_{\mathrm{M}}$ to achieve normally distributed estimates
for $\sigma^2$ and $\Omega$
more than make up for the additional cost of its Markov chain propagation.

\subsection{Configuration Selection}
\label{sec:sci}

%

The rapid progress in selective CI methods in recent years has greatly simplified
the selection of configurations for MSJ wave functions.
Although we expect that any modern selective CI method would work well with our approach,
we have taken advantage of existing links between the QMCPACK code \cite{KimQMCPACK2018}
and the CIPSI implementation within Quantum Package \cite{QuantumPackage1.1}
in order to extract configurations from the CIPSI variational wave function.
As studied by Dash et al, \cite{doi:10.1021/acs.jctc.8b00393}
a MSJ wave function can either be arrived at by
stopping the CIPSI algorithm once its expansion has reached the desired configuration
number, or by intentionally running CIPSI to a much larger configuration number
and then truncating to the number desired for use in MSJ.
Following their recommendation that the latter method is more effective, we have
for each of our systems iterated CIPSI with all non-core electrons and orbitals active
until each state's variational wave function contains at least 5,000 configurations,
after which we truncate to
the (typically much smaller) set of configurations used in a state's MSJ wave function
by retaining the configurations with the highest CIPSI weights for that state.
Although 5,000 configurations is far too few for even perturbatively-corrected
CIPSI to be converged for most of the systems we consider, the subsequent
addition of state-specific orbital optimization, Jastrow factors, and variance matching
allows this lightweight approach to be quite accurate.

\subsection{Variance Matching}
\label{sec:varMatch}

\begin{figure}[b]
    \centering
    \includegraphics[width=8.9cm]{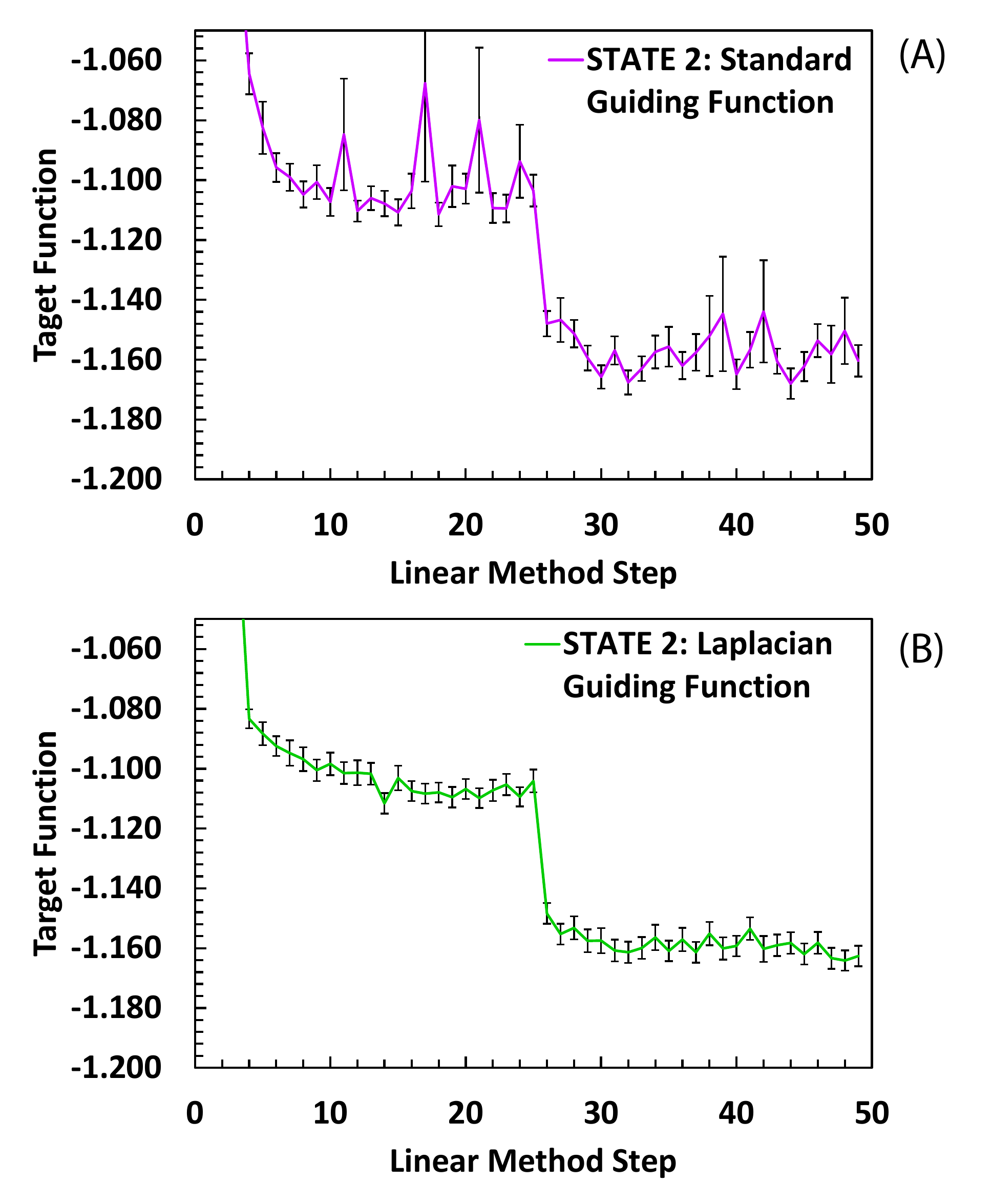}
    \caption{
      The target function $\Omega$ (in $(E_{\mathrm{h}})^{-1}$) during the
      optimization of the first singlet excited state of SCH$_{2}$ using a
      6 configuration wave function.
      Here we compare results using the traditional $|\Psi|^2$ guiding function
      (A) and our modified guiding function $|\Psi_{\mathrm{M}}|^2$ (B).
      The first 25 steps hold the orbitals fixed while optimizing the CI and Jastrow
      parameters while the last 25 steps optimize the orbitals as well.
    }
    \label{fig:sch2_guiding}
\end{figure}

In our previous work \cite{robinson2017varmatch} we showed that,
in practice, predictions of energy differences can be improved by
adjusting the sizes of different states' MSJ CI expansions such that
the states' energy variances $\sigma^2$ were equal.
The idea is to exploit the fact that $\sigma^2$ is essentially a
measurement for how close a state is to being a Hamiltonian eigenstate,
and, in the absence of a more direct measure of a states' energy error,
this measurement should be useful in ensuring that different states are
modeled at similar levels of quality so as to avoid bias.
That this approach helps improve cancellation of error is likely due,
at least in part, to the fact that the energies of low-lying states
tend to converge from above for large CI expansions, as the missing
tail of small-coefficient determinants means that what tends to be
missing is a full accounting of weak correlation effects, which in
low-lying states tend to lower a state's energy.

As before, we take the approach of evaluating both $E$ and $\sigma^2$
for a series of ground state wave functions of differing CI
expansion lengths so that we can interpolate to the expansion length
for which the ground state variance matches that of the excited state.
We perform the interpolation via the nonlinear fitting function (NLFF)
\begin{align}
    \label{eqn:nlff}
    f(N) = c + \frac{d}{N^{\alpha}}
\end{align}
where the functional form $f$ is used to interpolate both the
energy $E$ and energy variance $\sigma^2$ by fitting the values
$c$, $d$, and $\alpha$ for each case separately based on
an uncertainty-weighted least-squares fit.
Once these fits are made, we can estimate the expansion length $N$
for which the ground state variance would match that of the excited
state, and then, for that value of $N$, what we expect the ground
state energy would be.
Note that this is only one possible approach, as we could
equally well have fixed the ground state wave function and varied the
number of determinants in the excited state.
In some cases (e.g.\ see Section \ref{sec:BigCase}) it makes more sense
to seek an explicit match between two states' variances rather than
relying on interpolation.

\section{Results}

In the sections that follow, we will discuss results from a number of different
systems that we have tested our approach on.
Although we will point out the most relevant computational details as we go,
we refer readers to the Appendix for full details and geometries.
For software, we have implemented
our approach in a development version of QMCPACK \cite{KimQMCPACK2018}
and are working to ready the different components for inclusion in a future
public release.
We have also employed Molpro \cite{MOLPRO_paper}
for equation of motion coupled cluster with singles and doubles (EOM-CCSD),
complete active space self consistent field (CASSCF),
complete active space second order perturbation theory (CASPT2),
and Davidson-corrected multi-reference configuration interaction (MRCI+Q)
calculations, QChem \cite{shao:2015:qchem} for time-dependent
density functional theory (TD-DFT), Quantum Package for CIPSI calculations \cite{scemama_2015_30624}
, and Dice \cite{doi:10.1021/acs.jctc.6b00407,doi:10.1021/acs.jctc.6b01028}
for semi-stochastic heat bath CI (SHCI).

\subsection{Modified Guiding Function}

Before discussing our method's efficacy in predicting excitation energies, we would like
to emphasize the benefit of drawing samples from the modified guiding wave function
$\Psi_{\mathrm{M}}$.
As seen in Figure \ref{fig:sch2_guiding}, even the relatively simple optimization of
a 6-configuration wave function for the first excited singlet of thioformaldhyde
benefits significantly from the recovery of normal statistics for our estimates
of $\Omega$.
For a sample size of $N_s=768,000$ 
drawn from either $|\Psi_{\mathrm{M}}|^2$ or $|\Psi|^2$, the
worst uncertainties seen in $\Omega$ during the last 25 iterations
(as measured by a blocking analysis that assumes the statistics are normal)
are a factor of 4 smaller for the modified guiding case.
Even if $|\Psi|^2$ resulted in normal statistics (which it does not), this would
imply that our modified guiding function reduces the number of samples
needed to reach a given uncertainty by a factor of 16, which more than makes
up for the roughly 3 to 4 times increased cost per sample of propagating the
Markov chain for $|\Psi_{\mathrm{M}}|^2$.
It is also worth noting that, thanks to the decreased uncertainty, the
optimization that employed $|\Psi_{\mathrm{M}}|^2$ (which is used 
as the guiding function not
only when estimating $\Omega$ but also when evaluating the derivatives
needed by the linear method)
was able to converge to a lower average value of $\Omega$.
For the computational details for thioformaldhyde, see Section \ref{sec:sch2}.

%

\subsection{Simple Orbital Optimization Tests}

%
%
%

\begin{figure}[b]
    \centering
    \includegraphics[width=9.255cm]{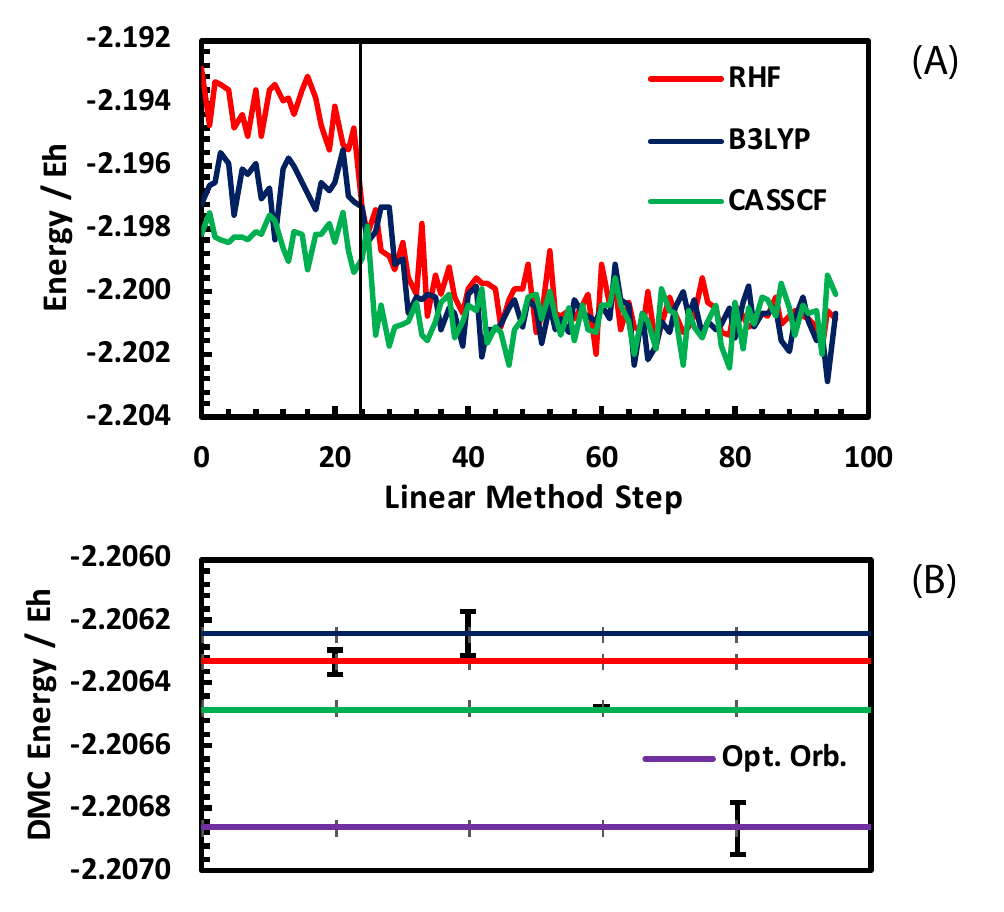}
    \caption{
      A: Energy vs optimization step for H$_{4}$ starting from different
      initial guesses for the orbitals.
      The orbitals are held fixed and only the CI coefficients and Jastrow variables
      optimized during the first 25 steps, after which all variables are optimized together.
      B: Time-step extrapolated DMC energies using the optimized trials wave functions
      before and after orbital optimization.
    }
    \label{fig:H4_robust}
\end{figure}

As an initial test of our orbital optimization implementation, we have attempted
to verify that it can remove the wave function's dependence on the initial
orbital basis in a number of simple test cases.
We begin with a low-symmetry, strongly correlated arrangement of four hydrogen
atoms (see Appendix for details) in which we construct a MSJ wave function for
the ground state using the 10 most important configurations from a ground state CASSCF (4e,10o) calculation.
With these configurations, we construct three different MSJ wave functions
by employing molecular orbitals from RHF, B3LYP, and the mentioned (4e,10o) CASSCF calculations.
As seen in Figure \ref{fig:H4_robust}A,
ground state energy optimizations in which the orbitals are held fixed and only
the Jastrow and CI coefficients varied result in three distinct energies, but
when we then optimize the orbitals as well all three wave functions converge
to the same energy, showing that the orbital optimization successfully removes
the dependence on starting orbitals in this case.
Although the effects are not large, the orbital optimization does have a
statistically significant effect on DMC energies, as Figure \ref{fig:H4_robust}B
shows the orbital-optimized nodal surface to be superior to that of any of
the three wave functions in which only the Jastrow and CI coefficients
were optimized.


\begin{figure*}[t]
    \centering
	\includegraphics[width=14.51cm]{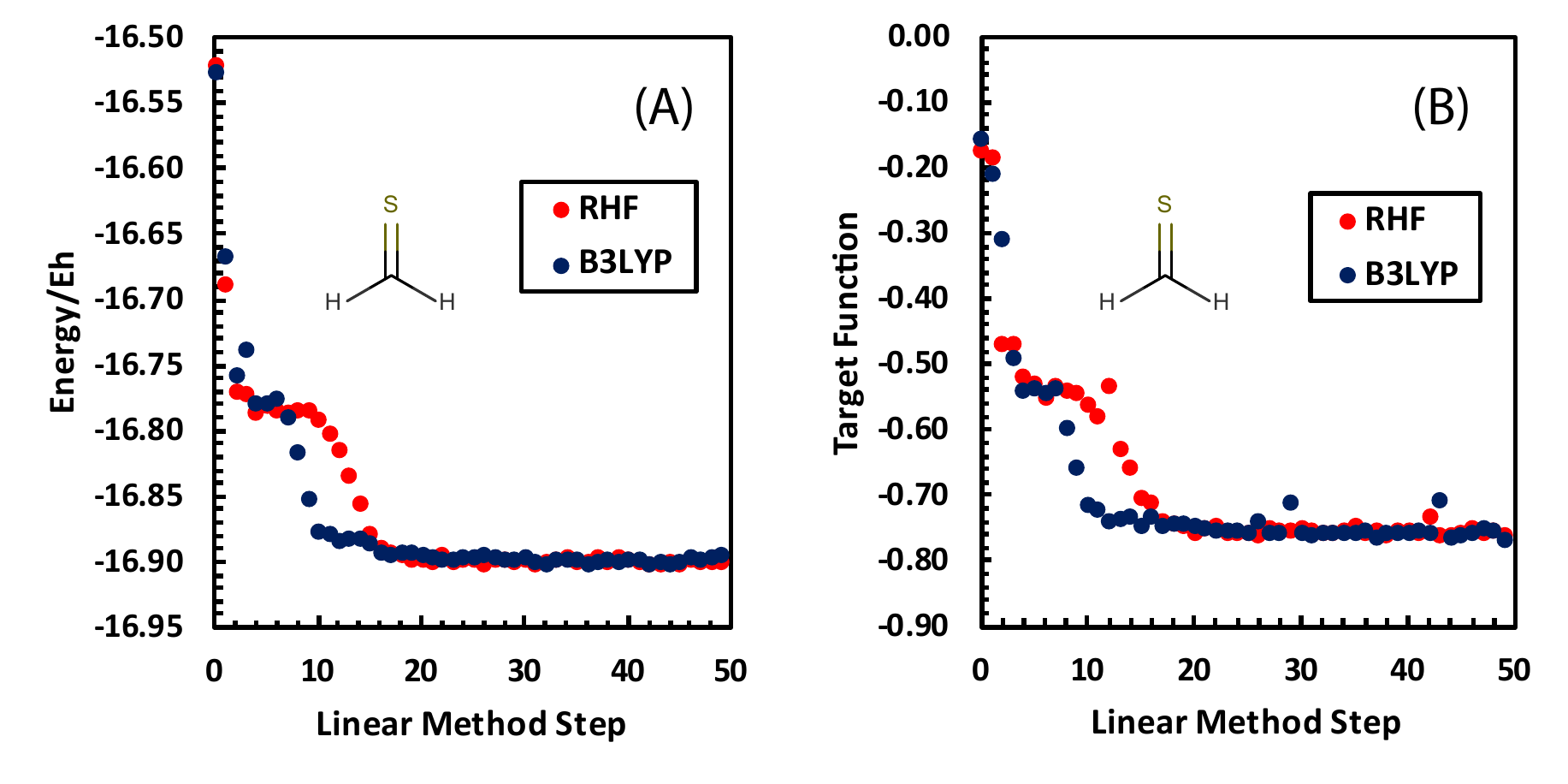}
    \caption{
    These plots represent how the energy $E$ (A) and target function $\Omega$ (B) change throughout the parameter optimization of two different MSJ wave functions which both represent the lowest singlet excited state of thioformaldehyde. These wave functions only differ in terms of the starting set of orbitals used, with the red points in subfigures (A) and (B) correspond to a wave function that begins with Hartree Fock orbitals while the blue points correspond to a wave function that begins with DFT B3LYP orbitals.
    }
    \vspace{4mm}
    \label{fig:sch2_robust}
    \includegraphics[width=14.51cm,keepaspectratio]{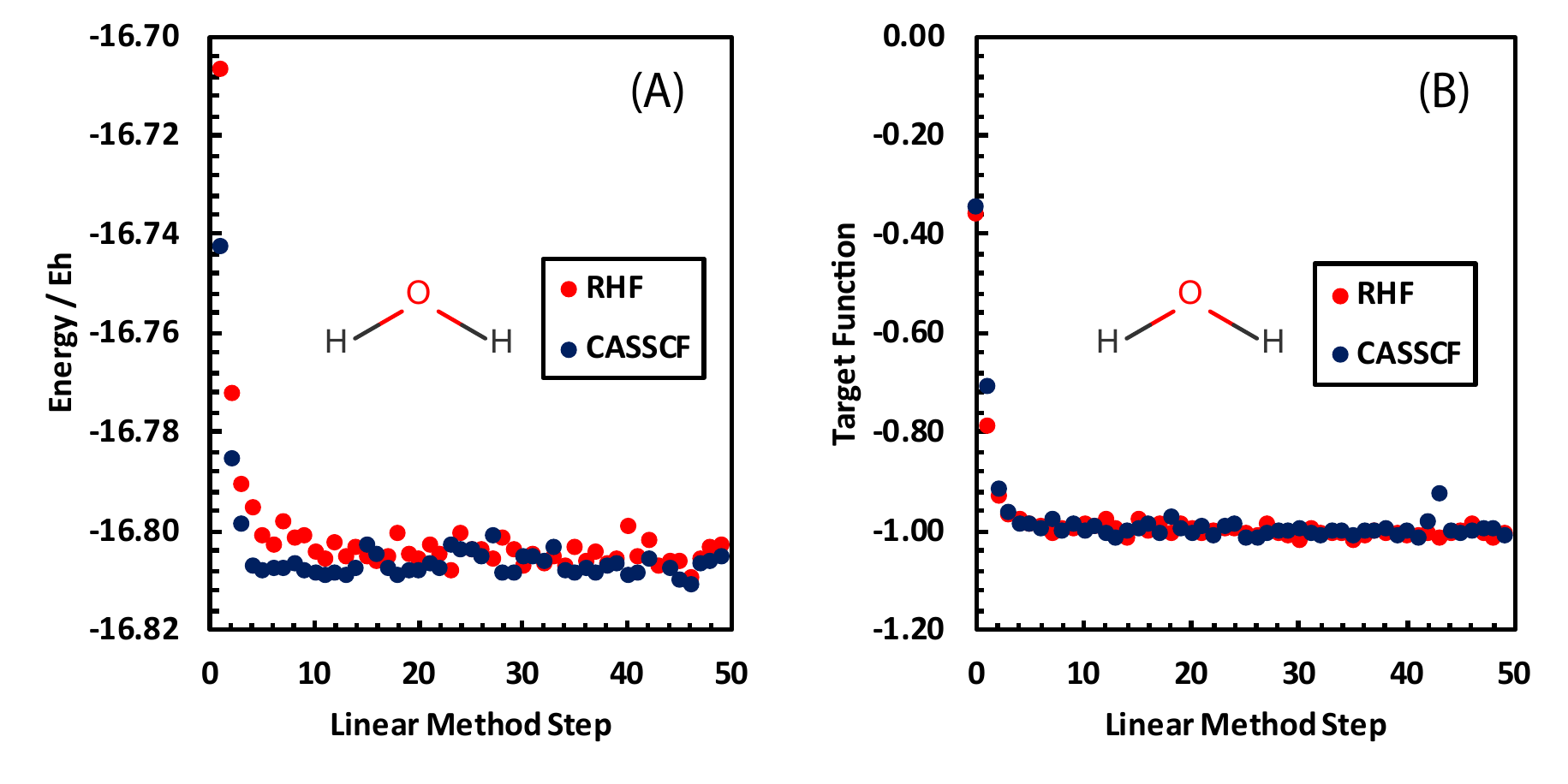}
    \caption{
    These plots represent how the energy $E$ (A) and target function $\Omega$ (B) change throughout the parameter optimization of two different MSJ wave functions which both represent the lowest singlet excited state of water. These wave functions only differ in terms of the starting set of orbitals used, with the red points in subfigures (A) and (B) correspond to a wave function that begins with Hartree Fock orbitals while the blue points correspond to a wave function that begins with state averaged CASSCF orbitals.}
    \label{fig:h2o_robust}
\end{figure*}

Moving on to thioformaldehyde and water, we find that the orbital optimization can
also remove starting orbital dependence in the excited state when variationally
minimizing $\Omega$.
In Figure \ref{fig:sch2_robust}, we see that in thioformaldehyde's
first excited singlet state,
a minimal 2 determinant MSJ wave function that captures the basic
open shell singlet structure optimizes to the same energy when
starting from either RHF or B3LYP orbitals.
Likewise, Figure \ref{fig:h2o_robust} shows the same behavior in water's
first singlet excited state when working with a 100-configuration MSJ
expansion and starting from either RHF orbitals or the orbitals
from an equally weighted two-state state-average (8e,8o)CASSCF.
While these systems can of course be treated accurately by other methods,
these tests are noteworthy as they represent the first excited state
calculations in which QMC removes a MSJ wave function's dependence
on the starting orbitals.

%
%

\subsection{Formaldimine}
\label{sec:formaldimine}

Photoisomerization is an important phenomenon that is responsible for many
interesting chemical events both in nature and the laboratory setting
and is an excellent example of the intersection between multi-reference
wave functions and excited states.
Molecules that undergo photoisomerization include rhodopsins, retinal proteins
involved in the conversion of light to electrical signals \cite{MazzoliniE2715},
and azobenzene, the prototypical photo switch studied for potential applications
as a molecular motor. \cite{doi:10.1021/cr970155y}
One of the smallest molecules that undergoes photoisomerization is formaldimine
(CH$_{2}$NH), in which the process proceeds following an absorption that
promotes it to its lowest singlet excited state.
\cite{DASILVA20111, dugave2006cis-trans, doi:10.1063/1.1777212}
The subsequent rotation around the C=N bond mixes the $\sigma$ and
$\pi$ orbitals and has been well studied,
including by molecular dynamics simulations,
\cite{doi:10.1063/1.475804,doi:10.1080/00268970512331339378}
and so this molecule makes for an excellent system in which to test the
effects of state-specific orbital optimization and variance matching
with modest MSJ expansions.

\begin{figure*}[t]
    \centering
    \includegraphics[width=14.51cm]{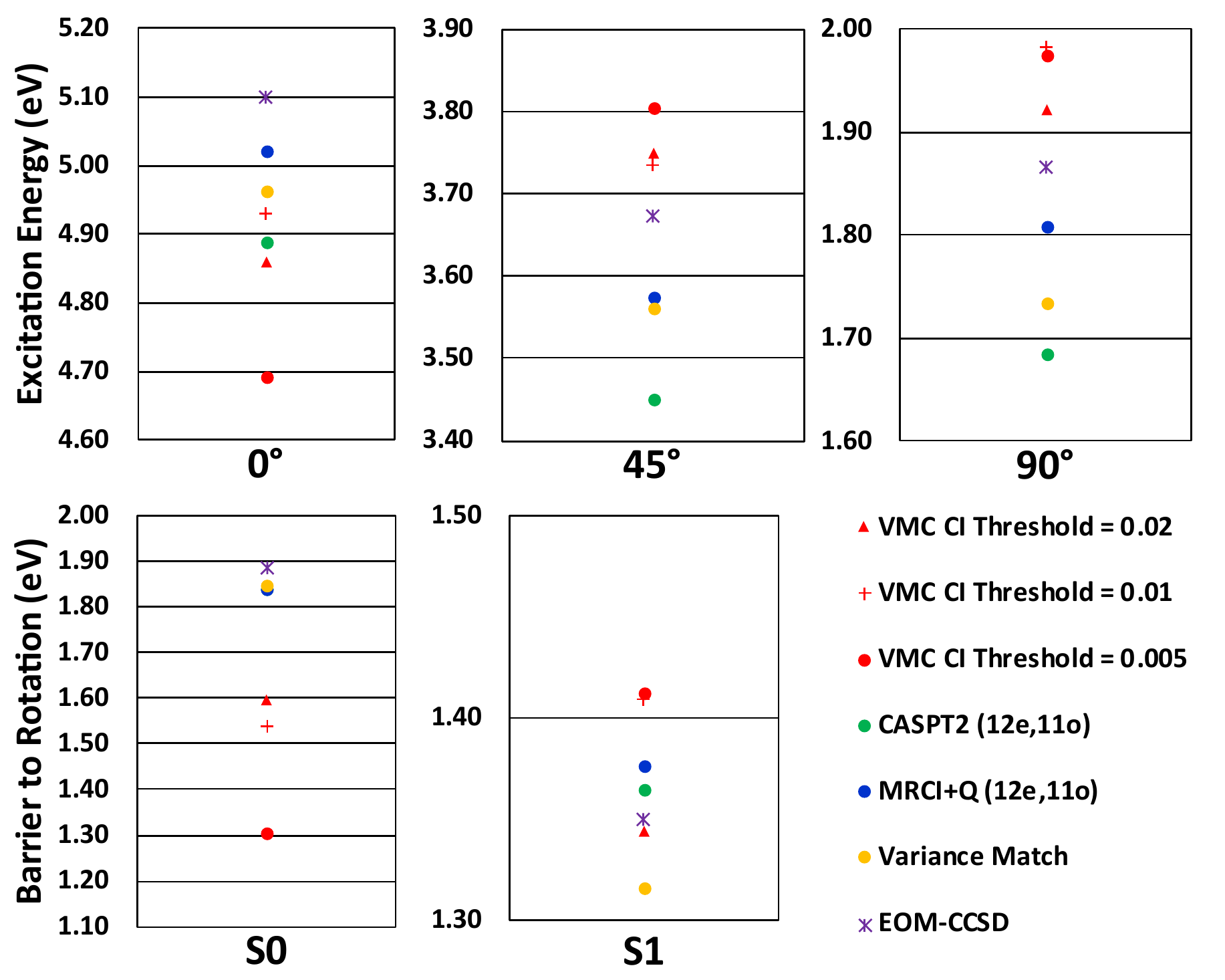}
    \caption{
  Tow row: excitation energies for CH$_{2}$NH at various torsion angles.
  Bottom row: barriers to rotation on the ground state (S0) and excited state (S1)
  surfaces. Statistical uncertainties were less than 0.04 ev in all cases
}
    \label{fig:CH3N_figure}
\end{figure*}

We model the ground and HOMO-LUMO ($n \rightarrow \pi^{*}$) excitation
using BFD effective core potentials and their VTZ basis
for torsion angles of 0, 45, and 90 degrees.
At each geometry, we start by optimizing the ground state
using 5 configurations for the 0 degree torsion coordinate geometry,
74 configurations for the 45 degree torsion coordinate geometry,
and 122 configurations for the 90 degree torsion coordinate geometry.
The configurations at each geometry were chosen by truncating the ground
state wave function from a CIPSI calculation in the RHF orbital basis.
We then optimize a series of excited state MSJ wave functions with approximately
$30$, $100$, and $600$ configurations taken from the same CIPSI
calculation with RHF orbitals and perform variance matching by applying
our NLFF to the excited states
(note this is the reverse of the single-excited-state/many-ground-state
procedure described above).
For determining the rotational barrier heights, we took energy differences
between the $90^\circ$ and $0^\circ$ geometries for both the ground S0
and excited S1 state.
In this case, variance matching was performed separately for the
S0 and S1 state barriers.
For each state, expansions of approximately $30$, $100$, and $600$ configurations
were used at the 90$^\circ$ geometry,
with which NLFFs were constructed to match the variances associated
with 4000 configurations for the $0^{o}$ S0 state and 649 configurations
for the $0^{o}$ S1 state.
In all cases, the configurations were selected from a truncated 5000
configuration CIPSI wave function.

One interesting observation here is that the CI expansions derived from
truncated CIPSI had as many as 57\% of their configurations lying
outside of a full valence (12e,12o) active space, which echoes previous
studies in which selective CI wave functions often find
many out-of-active-space configurations that prove to be more
important than most of the active space configurations.
That more than half of the $600$ most important determinants
in a CIPSI wave function lie outside of an active space
that contains 853,776 determinants reminds us how significant this
effect can be.
It is also important to note that, thanks to starting from a
truncated CIPSI wave function, our approach does not require selecting
an active space in this molecule, which removes one of the more
vexing difficulties of many multi-reference methods.
In general this approach might be expected to lead to unaffordable
selective CI calculations, but since we are truncating these
wave functions to very modest CI expansion lengths anyways, we do not
necessarily need CIPSI to converge to form a useful MSJ wave function
out of it, a point to which we will return in Section \ref{sec:BigCase}.


As seen in Figure \ref{fig:CH3N_figure}, our
approach --- which incorporates modest CI expansions, 
state-specific orbital optimization, and variance matching ---
predicts energy differences for excitations and barrier heights that
are within 0.1 eV of full-valence active space (12e,12o) state-averaged
MRCI+Q in all cases.
It is especially noteworthy that the alternative approach of taking energy
differences between fully optimized MSJ wave functions in which
configurations for both states are selected via a shared
CI coefficient threshold is much less reliable than the
variance matching approach.
This juxtaposition is a reminder that balancing wave function quality
is crucial when working with unconverged CI expansions.
Although this system is of course small enough that
large brute force expansions are feasible
(indeed CASPT2 also gives highly accurate results when used with
a full-valence CAS), we emphasize
that in large systems such an exhaustive approach will not be
feasible and CI-based methods will be forced to work with
incomplete CI expansions if they are to be used at all.
The fact that our overall approach is able to be successful in this case
despite using incomplete CI expansions is thus quite encouraging.

\subsection{Thioformaldehyde}
\label{sec:sch2}


\begin{table}[b]
\centering
\caption{
Excitation energies for the lowest singlet excitation in thioformaldehyde.
CASSCF, CASPT2, and MRCI+Q used a full-valence (12e,10o) active space, while SHCI
(and the CIPSI calculation from which we generated the MSJ expansion)
was performed for all 12 electrons in all orbitals.
}
\label{table:sch2}
\begin{tabular}{ll}
\hline \hline
Method              & h$\nu$ / eV  \\
\hline
2-state-SA-CASSCF       &\ 2.68    \\
SS-CASSCF               &\ 2.65    \\
2-state-SA-CASPT2       &\ 2.16    \\
SS-CASPT2 \hspace{25mm} &\ 2.13    \\
2-state-SA-MRCI+Q       &\ 2.31    \\
SS-MRCI+Q               &\ 2.32    \\
SHCI                    &\ 2.31(1) \\
EOM-CCSD    	       	&\ 2.40    \\
Variance Matched VMC    &\ 2.07(2) \\
\hline \hline
\end{tabular}
\end{table}

Although CH$_2$S undergoes similar chemical reactions as CH$_2$O
\cite{doi:10.1002/pol.1965.110030616,doi:10.1146/annurev.pc.34.100183.000335}
and sees a similar change (about 0.8 Debeye) in its MRCI dipole moment during its low-lying
$n \rightarrow \pi^{*}$ singlet excitation,
this absorption band is red-shifted so that what lay in the near ultraviolet in CH$_2$O
lies in the visible region \cite{doi:10.1146/annurev.pc.34.100183.000335} for CH$_2$S.
Given sulphur's more labile valence electrons and the persistence of modest
charge transfer character, CH$_2$S makes for an interesting test case, especially because
exact results can be benchmarked against even in a
triple zeta basis by employing large-core pseudopotentials so that only
12 electrons need to be simulated explicitly.

We took the approach described in the theory sections in order to try to
ensure balanced MSJ descriptions of the two states.
Specifically, we optimized the orbitals, CI coefficients, and Jastrow variables
for an 875 determinant MSJ expansion for the first excited singlet state using determinants
drawn from a two-state CIPSI calculation in the RHF orbital basis.
We then performed a series of analogous ground state optimizations using
expansions with 2, 5, 10, 50, 100, and 875 determinants taken from the same CIPSI
calculation.
After fitting our NLFF to the resulting energies and variances, our
variance-matched approach predicts an excitation energy similar to
that of full-valence state-specific CASPT2, as seen in Table \ref{table:sch2}.
The correct excitation energy in this basis --- confirmed by the agreement of
MRCI+Q with extrapolated SHCI --- is about 0.2 eV higher.
We therefore see that, although our approach to balancing the
accuracies of the different states' descriptions is not perfect,
it is able to provide reasonably high accuracy with very short CI expansions.

A feature of CH$_2$S that is worth noting is that our multi-reference quantum chemistry
results are quite insensitive to whether we a) ignore the molecule's
symmetry and arrive at the ground and first excited singlet states via a 2-state
state average or b) exploit the molecule's symmetry to do two ground state
calculations in different representations.
Table \ref{table:sch2} shows that the CASSCF, CASPT2, and MRCI+Q excitation energies
are little changed when we switch between these state-averaged and state-specific approaches.
Certainly our ability to afford a full-valence CAS in thioformaldehyde contributes to this
insensitivity, but in any case our MSJ approach's ability to tailor the orbitals in a
state-specific manner is clearly not essential here.
We now turn to a case in which state averaging is more problematic
in order to emphasize the advantages of
a fully variational approach with state-specific orbitals.

\subsection{$\big[\hspace{0.5mm}$C$_{3}$N$_{2}$O$_{2}$H$_{4}$Cl$\hspace{0.5mm}\big]^{-}$}
\label{sec:BigCase}

\begin{figure}[b]
    \centering
    \includegraphics[width=4.5cm]{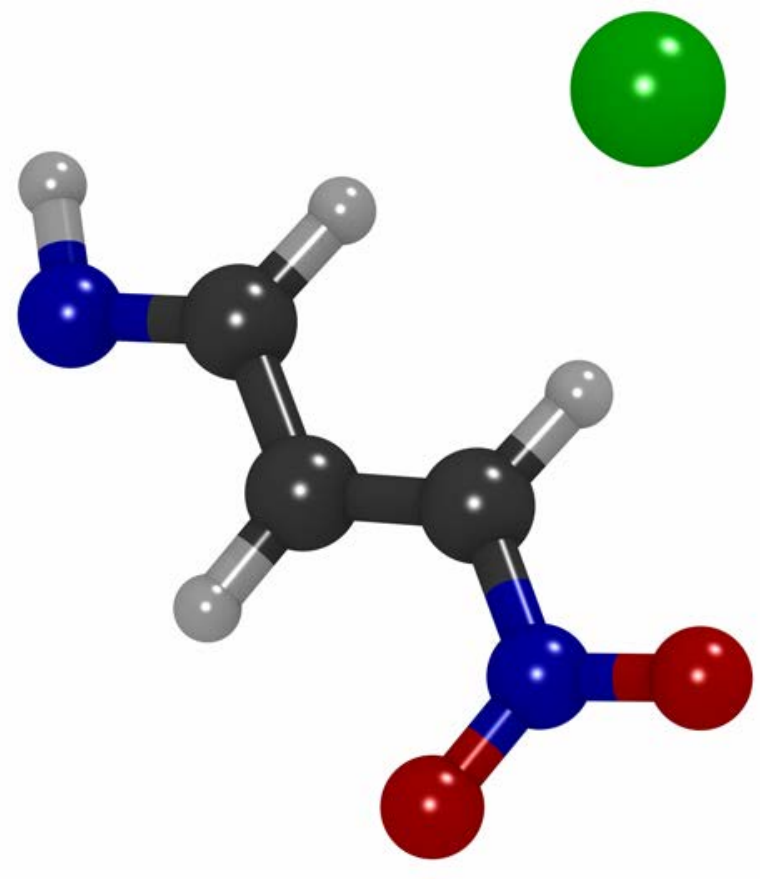}
    \caption{
    The $\big[\hspace{0.5mm}$C$_{3}$N$_{2}$O$_{2}$H$_{4}$Cl$\hspace{0.5mm}\big]^{-}$
    anion used in our chlorine-to-$\pi^{*}$ charge transfer example.
    In the ground state, the charge is localized on the (green) chlorine atom,
    while many of the excited states have the charge distributed in
    the $\pi$ network.
    }
    \label{fig:big_molecule}
\end{figure}

Compared to CH$_2$S, many of the low-lying excited states of
$\big[\hspace{0.5mm}$C$_{3}$N$_{2}$O$_{2}$H$_{4}$Cl$\hspace{0.5mm}\big]^{-}$,
shown in Figure \ref{fig:big_molecule}, have very strong charge transfer character.
In the ground state, this system localizes the extra electron on the Cl atom,
which although not bonded covalently to the main molecule is attracted
by dipole/charge interactions and dispersion forces (we determined its location through
MP2/cc-pVDZ geometry optimization).
According to EOM-CCSD, the first four singlet excited states all transfer
an electron into the lowest $\pi^{*}$ orbital.
In order of increasing excitation energy, these transfers come from
the two Cl in-plane p orbitals (3.56 and 3.74 eV),
the Cl out-of-plane p orbital (3.86 eV),
and an oxygen in-plane p orbital (3.91 eV).
In addition, there are multiple other $n$$\rightarrow$$\pi^*$ and
$\pi$$\rightarrow$$\pi^*$ singlet transitions in the 4$-$7 eV range.
Although some of these states have strong charge transfer character,
EOM-CCSD is a good reference for their excitation energies thanks to:
(a) the fact that they are all singly excited states and (b) EOM-CCSD's doubles
operator's ability to provide the state-specific orbital relaxations
that are so crucial for charge transfer.

\begin{table}[t]
\centering
\caption{
  Excitation energies in eV for the totally symmetric chlorine-to-$\pi^*$
  transition in
  $\big[\hspace{0.5mm}$C$_{3}$N$_{2}$O$_{2}$H$_{4}$Cl$\hspace{0.5mm}\big]^{-}$
  using BFD effective core potentials and the corresponding VDZ basis set.
  \cite{doi:10.1063/1.2741534}
  CASSCF and CASPT2 results are based on either state averaged (SA)
  orbitals or nearly state specific (SS) orbitals.
  CASPT2 employed either Roos-Andersson (RA) level shifts \cite{ROOS1995215}
  or the ionization potential electron affinity (IPEA) approach \cite{GHIGO2004142}
  to deal with intruder states.
  See Section \ref{sec:BigCase} for details.
}
\label{table:c3cln2o2h4}
\begin{tabular}{llll}
\hline \hline
Method &
\multicolumn{2}{l}{\hspace{2mm}Level Shift}  &
\hspace{2mm}h$\nu$    \\
\hline    
4-state-SA-CASSCF  & \multicolumn{2}{l}{N/A}       & 2.127 \\ 
2-state-SA-CASSCF  & \multicolumn{2}{l}{N/A}       & 3.980 \\ 
(95/5)-SS-CASSCF   & \multicolumn{2}{l}{N/A}       & 4.669 \\ 
4-state-SA-CASPT2  & RA   & $\varepsilon$ = 0.2    & 3.369 \\ 
4-state-SA-CASPT2  & RA   & $\varepsilon$ = 0.3    & 3.340 \\ 
4-state-SA-CASPT2  & IPEA & $\varepsilon$ = 0.25   & 3.461 \\ 
2-state-SA-CASPT2  & RA   & $\varepsilon$ = 0.2    & 3.285 \\ 
2-state-SA-CASPT2  & RA   & $\varepsilon$ = 0.3    & 3.304 \\ 
2-state-SA-CASPT2  & IPEA & $\varepsilon$ = 0.25   & 3.546 \\ 
(95/5)-SS-CASPT2   & RA   & $\varepsilon$ = 0.2    & 3.365 \\ 
(95/5)-SS-CASPT2   & RA   & $\varepsilon$ = 0.3    & 3.387 \\ 
(95/5)-SS-CASPT2   & IPEA & $\varepsilon$ = 0.25   & 3.541 \\ 
PBE0   \hspace{31mm} & \multicolumn{2}{l}{N/A}	\hspace{20mm} & 1.233 \\
B3LYP	             & \multicolumn{2}{l}{N/A}			      & 0.903 \\
M06-2X	             & \multicolumn{2}{l}{N/A}			      & 0.903 \\
wB97X-V 	         & \multicolumn{2}{l}{N/A} 			      & 3.348 \\
wB97M-V 	         & \multicolumn{2}{l}{N/A} 			      & 3.187 \\
EOM-CCSD    	     & \multicolumn{2}{l}{N/A}    			  & 3.856 \\
Variance Matched VMC & \multicolumn{2}{l}{N/A}         	      & 3.80(3) \\
\hline \hline
\end{tabular}
\end{table}

In contrast, multi-reference methods are harder to
use effectively here, both due to the molecule's larger size
and due to the difficulties
that state averaging encounters when faced with states
that have large differences between their charge distributions.
Most notably, the ground state and any excited states that do not
involve the chlorine atom have dipoles that differ by more than
10 Debeye compared to excited states that have transferred an
electron from the chlorine into the $\pi$ network.
These differences mean that even orbitals outside the active space
are expected to relax significantly when transferring between these two
sets of states, making it exceedingly challenging to arrive at
a good set of state averaged orbitals that are appropriate for
all of the low-lying states.
This difficulty can be seen even if we restrict our attention to
the two lowest states in the totally symmetric representation,
which are the ground state (charge on the chlorine) and the
out-of-plane-Cl-3p$\hspace{0.8mm}\rightarrow\pi^*$ excitation.
As seen in Table \ref{table:c3cln2o2h4},
there is a 0.7 eV difference in the excitation energy
predicted by an equally-weighted 2-state state averaged CASSCF
calculation and a calculation in which we attempt to
optimize the orbitals state specifically by using
95\%/5\% and 5\%/95\% weightings
(note that our active space distributed ten electrons among
the three chlorine 3p orbitals
and the five $\pi$/$\pi^*$ orbitals closest to the gap).
The CASSCF sensitivity to state averaging is thus much
higher here than it was in CH$_2$S, making this system
an interesting test case for a multi-configurational
approach that can achieve fully state specific orbital optimization.

While the state-averaging sensitivity is already worrisome
here, we should point out that,
were this molecule not $C_s$ symmetric, predicting
this out-of-plane-Cl-3P$\hspace{0.8mm}\rightarrow\pi^*$
excitation energy via standard multi-reference
methods would be even more difficult.
Indeed, when we did not exploit the symmetry and instead treated
the molecule as if it were $C_1$, we were unable to find either a
2-state or 3-state state averaged CASSCF that contained the 
out-of-plane-Cl-3p$\hspace{0.8mm}\rightarrow\pi^*$ excitation,
despite trying numerous initial guesses and optimization methods.
Either one or both of the two in-plane-Cl-3p$\hspace{0.8mm}\rightarrow\pi^*$
excitations ended up being lower in energy,
or, for some initial guesses, the orbitals optimized to be so
favorable for states with ground-state-like charge distributions
that none of the excitations turned out to involve moving charge
away from the chlorine atom.
It was only when we resorted to a 4-state state average CASSCF
that we were able to find the desired state, which turned out
to come out alongside the ground state and the two
in-plane-Cl-3p$\hspace{0.8mm}\rightarrow\pi^*$ excitations.
In this case, three of the four states in the state average
have a dipole greatly different from that of the ground state,
and so as one would expect the orbital optimization favored them
at the expense of the ground state, resulting in a much too small
CASSCF excitation energy, as seen in Table \ref{table:c3cln2o2h4}.
As real chemical environments such as protein superstructures or
solvents typically remove symmetry, one realizes that
these types of state averaging difficulties will be quite common
when attempting to model charge transfer in large molecules
and realistic environments.

Of course, one should not expect quantitative accuracy from
CASSCF excitation energies whether or not there are state averaging
concerns, as these calculations omit the weak correlation effects
of orbitals and electrons outside the active space.
For larger molecules and active spaces, CASPT2 is much more
affordable than MRCI+Q, and so we have employed it here both
to include weak correlation effects and in the hope that it can
via its singles excitations help to put back the state-specific
orbital relaxations that are inevitably compromised during
state averaging.
Unfortunately, we found that the excited state we are after
suffers from intruder state \cite{ROOS1995215}
problems in all cases here, regardless of how the state-averaging
was handled, and so we were forced to employ level shifts in order
to avoid unphysically large perturbative corrections.
We found that for each of our state-averaging and state-specific
approaches, the value and type of level shift made a noticeable
difference in the CASPT2 excitation energies, which is not ideal.
Overall, the CASPT2 results's errors ranged from about 0.3 to 0.5 eV
when compared to EOM-CCSD,
which does leave something to be desired but is nonetheless
an improvement over TD-DFT.

For our MSJ treatment of this system, we began by iterating
the variational stage of a 4 state CISPI calculation
(ignoring symmetry) in the RHF orbital basis until
it had identified more than 5,000 important determinants.
At this point we found that the out-of-plane charge transfer
state we are focusing on was the fourth CIPSI root and we ended
the CIPSI iterations, even though for a system of this size the
expansion procedure was certainly far from converged.
This incomplete CIPSI does not present an issue for us, though,
as we imported only the fourth root's 100 most important determinants
into our MSJ wave function, and we expect
that in a 5,000-determinant expansion the importance-ordering
of the first few hundred should be well converged.
We then applied our variational excited state methodology
to optimize the orbitals, CI coefficients, and Jastrow variables
for this MSJ wave function and evaluated its variance.
Repeating this procedure for the ground state
(the first root from the same incomplete CIPSI expansion)
we found that the ground state MSJ wave function required
very few (roughly 10) determinants in order to have a variance
smaller than that of the excited state.
Given that the ground state variance will be a much more noticeably
discreet function of determinant number for such short expansions,
we decided to forgo the NLFF (which makes more sense when the
variance is changing close to continuously with determinant number)
and instead varied the number of ground state determinants by hand to
find the expansion whose variance most closely matched that
of the excited state.
This approach found that the 10-determinant ground state MSJ wave
function (with optimized orbitals, CI, and Jastrow) made for the
best match, which resulted in a predicted excitation energy within
0.1 eV of the EOM-CCSD benchmark, as seen in Table \ref{table:c3cln2o2h4}.
Thus, as in the smaller systems, we find that a combination
of short CIPSI-derived expansions, orbital optimization, and
variance matching delivers a reasonably high accuracy
even in a case whose size and strong charge transfer character
complicates the application of traditional multi-reference methods.

\section{Conclusions}
\label{sec:conclusion}

By combining efficient methods for multi-Slater-Jastrow expansions, state-specific orbital relaxations via
excited state variational principles, compact CI expansions generated by selective CI methods, and
variance matching to help balance the treatments of different states, we have shown that accuracies
on the order of a couple tenths of electron volts can be achieved when predicting excitation energies.
These findings are especially exciting in the context of larger molecules, where traditional high-accuracy
methods based on state averaging become challenging and the exponential scaling of selective CI 
approaches prevents them from being effective in isolation.
In future, the prospect of combining these state-specific multi-Slater wave functions with diffusion
Monte Carlo promises to allow multi-determinant wave functions from very large active spaces
(via unconverged selective CI) to be taken towards the basis set limit.
Indeed, the straightforward systematic improvability of the ansatz may well allow fully correlated
excitation energies to be converged in the basis set limit in molecules well beyond the reach of
even perturbatively-corrected selective CI methods.

Although in this study we have emphasized the importance of state-specific orbital relaxation for
charge transfer states, the potential benefits of the approach are much broader.
Double excitations, for example, are not treated accurately by EOM-CCSD, and so in systems where
state averaging is required for traditional multi-reference methods to be stable this new
state specific approach could offer a powerful alternative.
In the solid state, the application of these systematically improvable techniques to study
excitations in strongly correlated metal oxides is also quite promising, although methods for
generating the determinant expansion are less well developed in this area.
In practice, realizing the potential of the approach we have presented in larger systems
will require continuing improvements in VMC
wave function optimization methods in order to handle ever-larger sets of variational parameters.
Between these methodological priorities and the many exciting applications that are possible, we look forward
to a wide variety of future work with excited-state-specific multi-Slater Jastrow wave functions.


\section{Acknowledgements}

The authors thank Ksenia Bravaya for a helpful conversation.
This work was supported by the Office of Science,
Office of Basic Energy Sciences, the US Department of Energy,
Contract No. DE-AC02-05CH11231. SDPF acknowledges additional support from the Department of Energy's Stewardship Science Graduate Fellowship under grant No. DE-NA0003864.
Calculations were performed
using the Berkeley Research Computing Savio cluster
and the Argonne Leadership Computing Facility.
An award of computer time was provided by the Innovative
and Novel Computational Impact on Theory and Experiment
(INCITE) program. This research has used resources of the
Argonne Leadership Computing Facility, which is a DOE
Office of Science User Facility supported under Contract
No. DE-AC02-06CH11357. 

\clearpage

\section{Appendix: Computational Details}

For Quantum Monte Carlo calculations all wave function
parameters were optimized by variationally minimizing
$\Omega$ using the $\omega$-update scheme presented by
Shea and Neuscamman. \cite{Shea:2017:scesvp}
Both the two-body and all one-body Jastrow factors took
a Bspline form with a cutoff at 10 bohr, and using 10 spline points.
\cite{KimQMCPACK2018}
\subsection{H$_{4}$}

Our skew arrangement of four H atoms was chosen to remove all symmetry
and to create a simple, small system in which strong correlation was
present so that orbital relaxations in a small MSJ expansion would
be expected to make a difference.
We employed BFD effective core potentials and the corresponding VTZ basis
\cite{doi:10.1063/1.2741534}
and placed the atoms at the positions given below in Angstroms.
The configurations for the MSJ were chosen by using the 10 most important configurations from a ground state CASSCF (4e,10o) calculation.
\begin{table}[h]
\begin{tabular}{llll}
H & $\quad$ 0.0000000000 & $\quad$ 0.0000000000 & $\quad$ 0.0000000000 \\
H & $\quad$ 1.8897259877 & $\quad$ 0.0000000000 & $\quad$ 0.0000000000 \\
H & $\quad$ 0.0000000000 & $\quad$ 0.0000000000 & $\quad$ 2.8345889816 \\
H & $\quad$ 0.0000000000 & $\quad$ 0.0000000000 & $\quad$ 5.6691779632
\end{tabular}
\end{table}

\subsection{H$_{2}$O}

For our calculations on water we employed
BFD effective core potentials with the VDZ basis \cite{doi:10.1063/1.2741534}
at the experimental equilbirum geometry \cite{hoy1979precise}
given in Angstroms below.
The 100 configurations for the MSJ excited state singlet
were chosen as the largest-weight configurations in
the excited state of
a two-state full-valence CIPSI calculation in the RHF orbital basis.
\begin{table}[h]
\begin{tabular}{l r@{.}l r@{.}l r@{.}l}
O & $\quad$ 0&0000 & $\quad$  0&0000 & $\quad$  0&1173 \\
H & $\quad$ 0&0000 & $\quad$  0&7572 & $\quad$ -0&4692 \\
H & $\quad$ 0&0000 & $\quad$ -0&7572 & $\quad$ -0&4692
\end{tabular}
\end{table}

\subsection{CH$_2$S}

For thioformaldehyde we used  
BFD effective core potentials with their VTZ basis
\cite{doi:10.1063/1.2741534}
and the following geometry, in Angstroms.
\begin{table}[h]
\begin{tabular}{l r@{.}l r@{.}l r@{.}l}
S & $\quad$ -4&9615006425 & $\quad$ 2&6553412397 & $\quad$  0&0000217073 \\
C & $\quad$ -4&9017991394 & $\quad$ 1&0716634201 & $\quad$ -0&0001062888 \\
H & $\quad$ -5&5890742022 & $\quad$ 0&4742274771 & $\quad$  0&5685871400 \\
H & $\quad$ -4&1719760160 & $\quad$ 0&5275278631 & $\quad$ -0&5685025585
\end{tabular}
\end{table}

\hphantom{Hi there.}

\subsection{$\big[\hspace{0.5mm}$C$_{3}$N$_{2}$O$_{2}$H$_{4}$Cl$\hspace{0.5mm}\big]^{-}$}

For our chlorine-to-$\pi^{*}$ charge transfer system,
we used the geometry given in Angstroms below that was
arrived at via an MP2/cc-pVDZ geometry optimization in Molpro
for the closed-shell anionic ground state.
For all excitation energy evaluations, we employed
BFD effective core potentials
with the corresponding VDZ basis. \cite{doi:10.1063/1.2741534}
\begin{table}[h]
\begin{tabular}{l r@{.}l r@{.}l r@{.}l}
N  & $\quad$ -2&9058516510 & $\quad$ 0&0000000000 & $\quad$ -1&4601212300 \\
C  & $\quad$ -1&6406693060 & $\quad$ 0&0000000000 & $\quad$ -1&1581565780 \\
C  & $\quad$ -1&2646457880 & $\quad$ 0&0000000000 & $\quad$  0&2633575220 \\
C  & $\quad$  0&0390698630 & $\quad$ 0&0000000000 & $\quad$  0&6172500750 \\
N  & $\quad$  0&4054577830 & $\quad$ 0&0000000000 & $\quad$  2&0353753730 \\
O  & $\quad$  1&6182721380 & $\quad$ 0&0000000000 & $\quad$  2&2841014810 \\
O  & $\quad$ -0&4846283820 & $\quad$ 0&0000000000 & $\quad$  2&8946943050 \\
Cl & $\quad$  1&5876389750 & $\quad$ 0&0000000000 & $\quad$ -2&3723480080 \\
H  & $\quad$  0&8777584070 & $\quad$ 0&0000000000 & $\quad$ -0&1013594360 \\
H  & $\quad$ -2&0565965790 & $\quad$ 0&0000000000 & $\quad$  1&0201589650 \\
H  & $\quad$ -0&7953016620 & $\quad$ 0&0000000000 & $\quad$ -1&8742706390 \\
H  & $\quad$ -2&9621125330 & $\quad$ 0&0000000000 & $\quad$ -2&4917011270
\end{tabular}
\end{table}


\subsection{CH$_{2}$NH}

For formaldimine, we used 
BFD effective core potentials and their VTZ basis. \cite{doi:10.1063/1.2741534}
All active space methods were based on equally-weighted
two-state state averaged CASSCF wave functions. A (2e,2o) active space ground state CASSCF geometry optimization was performed with the constraint that the dihedral angle of the molecule remain fixed at 0, 45, or 90 degree.

\begin{table}[h]
\begin{tabular}{l r@{.}l r@{.}l r@{.}l}
\multicolumn{7}{c}{Torsion Coordinate 0$^{\circ}$} \\[1mm]
C & $\quad$ 0&0000000000 & $\quad$  0&0222705613 & $\quad$ -0&6225060885 \\
N & $\quad$ 0&0000000000 & $\quad$ -0&0820734375 & $\quad$  0&6148349963 \\
H & $\quad$ 0&0000000000 & $\quad$  0&9991273614 & $\quad$ -1&0742328325 \\
H & $\quad$ 0&0000000000 & $\quad$ -0&8651411291 & $\quad$ -1&2314451255 \\
H & $\quad$ 0&0000000000 & $\quad$  0&7411514791 & $\quad$  1&1797292763 \\[4mm]
%
%
\multicolumn{7}{c}{Torsion Coordinate 45$^{\circ}$} \\[1mm]
C & $\quad$ -0&0079599003 & $\quad$  0&0166001095 & $\quad$ -0&6325884466 \\
N & $\quad$  0&0416062329 & $\quad$ -0&0648316530 & $\quad$  0&6270535275 \\
H & $\quad$  0&1330377950 & $\quad$  0&9587507613 & $\quad$ -1&1388276756 \\
H & $\quad$ -0&1929029663 & $\quad$ -0&8533530806 & $\quad$ -1&2431487361 \\
H & $\quad$ -0&4234569109 & $\quad$  0&5977132175 & $\quad$  1&2063798375 \\[4mm]
%
%
\multicolumn{7}{c}{Torsion Coordinate 90$^{\circ}$} \\[1mm]
C & $\quad$ -0&0144604887 & $\quad$  0&0000000000 & $\quad$ -0&6239939985 \\
N & $\quad$  0&0700010603 & $\quad$  0&0000000000 & $\quad$  0&6267505389 \\
H & $\quad$ -0&0579593274 & $\quad$  0&8804123024 & $\quad$ -1&2681443330 \\
H & $\quad$ -0&0579593274 & $\quad$ -0&8804123024 & $\quad$ -1&2681443330 \\
H & $\quad$ -0&6845247458 & $\quad$  0&0000000000 & $\quad$  1&2624877883
\end{tabular}
\end{table}

\bibliographystyle{aip}

\begin{thebibliography}{10}

\bibitem{doi:10.1002/anie.200804709}
A.~Mishra, M.~Fischer, and P.~Bäuerle,
\newblock Angew. Chem. Int. Ed. {\bf 48}, 2474 (2009).

\bibitem{doi:10.1021/cr900182s}
Y.-J. Cheng, S.-H. Yang, and C.-S. Hsu,
\newblock Chem. Rev. {\bf 109}, 5868 (2009),
\newblock PMID: 19785455.

\bibitem{C2CS35266D}
F.~E. Osterloh,
\newblock Chem. Soc. Rev. {\bf 42}, 2294 (2013).

\bibitem{doi:10.1021/nn9015423}
S.~C. Roy, O.~K. Varghese, M.~Paulose, and C.~A. Grimes,
\newblock ACS Nano {\bf 4}, 1259 (2010),
\newblock PMID: 20141175.

\bibitem{doi:10.1021/ar900209b}
D.~Gust, T.~A. Moore, and A.~L. Moore,
\newblock Acc. Chem. Res. {\bf 42}, 1890 (2009),
\newblock PMID: 19902921.

\bibitem{subotnik2011cis_ct}
J.~E. Subotnik,
\newblock J. Chem. Phys. {\bf 135}, 071104 (2011).

\bibitem{chaiSystematicLRCDFT}
J.-D. Chai and M.~Head-Gordon,
\newblock J. Chem. Phys. {\bf 128}, 084106 (2008).

\bibitem{Krylov:2008:eom_cc_review}
A.~I. Krylov,
\newblock Annu. Rev. Phys. Chem. {\bf 59}, 433 (2008).

\bibitem{gilbert2008mom}
A.~T.~B. Gilbert, N.~A. Besley, and P.~M.~W. Gill,
\newblock J. Phys. Chem. A {\bf 112}, 13164 (2008).

\bibitem{umrigar2007alleviation}
C.~Umrigar, J.~Toulouse, C.~Filippi, S.~Sorella, and R.~G. Hennig,
\newblock Phys. Rev. Lett. {\bf 98}, 110201 (2007).

\bibitem{giner2013using}
E.~Giner, A.~Scemama, and M.~Caffarel,
\newblock Can. J. Chem. {\bf 91}, 879 (2013).

\bibitem{scemama2014accurate}
A.~Scemama, T.~Applencourt, E.~Giner, and M.~Caffarel,
\newblock J. Chem. Phys. {\bf 141}, 244110 (2014).

\bibitem{giner2015fixed}
E.~Giner, A.~Scemama, and M.~Caffarel,
\newblock J. Chem. Phys. {\bf 142}, 044115 (2015).

\bibitem{doi:10.1063/1.1777212}
F.~Schautz, F.~Buda, and C.~Filippi,
\newblock J. Chem. Phys. {\bf 121}, 5836 (2004).

\bibitem{caffarel2016communication}
M.~Caffarel, T.~Applencourt, E.~Giner, and A.~Scemama,
\newblock J. Chem. Phys. {\bf 144}, 151103 (2016).

\bibitem{LoosSCIDMC2018}
A.~Scemama, A.~Benali, D.~Jacquemin, M.~Caffarel, and P.-F. Loos,
\newblock J. Chem. Phys. {\bf 149}, 034108 (2018).

\bibitem{doi:10.1021/acs.jctc.8b00393}
M.~Dash, S.~Moroni, A.~Scemama, and C.~Filippi,
\newblock J. Chem. Theory Comput. {\bf 14}, 4176 (2018).

\bibitem{foulkes2001quantum}
W.~Foulkes, L.~Mitas, R.~Needs, and G.~Rajagopal,
\newblock Rev. Mod. Phys. {\bf 73}, 33 (2001).

\bibitem{Zhao:2016:dir_tar}
L.~Zhao and E.~Neuscamman,
\newblock J. Chem. Theory Comput. {\bf 12}, 3436 (2016).

\bibitem{robinson2017varmatch}
P.~J. Robinson, S.~D. Pineda~Flores, and E.~Neuscamman,
\newblock J. Chem. Phys. {\bf 147}, 164114 (2017).

\bibitem{zhao2017blocked}
L.~Zhao and E.~Neuscamman,
\newblock J. Chem. Theory Comput. {\bf 13}, 2604 (2017).

\bibitem{Shea:2017:scesvp}
J.~A.~R. Shea and E.~Neuscamman,
\newblock J. Chem. Theory Comput. {\bf 13}, 6078 (2017).

\bibitem{schriber2016communication}
J.~Schriber and F.~Evangelista,
\newblock J. Chem. Phys. {\bf 144}, 161106 (2016).

\bibitem{doi:10.1063/1.4955109}
N.~M. Tubman, J.~Lee, T.~Y. Takeshita, M.~Head-Gordon, and K.~B. Whaley,
\newblock J. Chem. Phys. {\bf 145}, 044112 (2016).

\bibitem{doi:10.1021/acs.jctc.6b00407}
A.~A. Holmes, N.~M. Tubman, and C.~J. Umrigar,
\newblock J. Chem. Theory Comput. {\bf 12}, 3674 (2016),
\newblock PMID: 27428771.

\bibitem{doi:10.1021/acs.jctc.6b01028}
S.~Sharma, A.~A. Holmes, G.~Jeanmairet, A.~Alavi, and C.~J. Umrigar,
\newblock J. Chem. Theory Comput. {\bf 13}, 1595 (2017),
\newblock PMID: 28263594.

\bibitem{garniron2018selected}
Y.~Garniron, A.~Scemama, E.~Giner, M.~Caffarel, and P.-F. Loos,
\newblock J. Chem. Phys. {\bf 149}, 064103 (2018).

\bibitem{table_method}
B.~K. Clark, M.~A. Morales, J.~McMinis, J.~Kim, and G.~E. Scuseria,
\newblock J. Chem. Phys. {\bf 135}, 244105 (2011).

\bibitem{table_deriv}
C.~Filippi, R.~Assaraf, and S.~Moroni,
\newblock J. Chem. Phys. {\bf 144}, 194105 (2016).

\bibitem{table_deriv2}
R.~Assaraf, S.~Moroni, and C.~Filippi,
\newblock J. Chem. Theory Comput. {\bf 13}, 5273 (2017),
\newblock PMID: 28873307.

\bibitem{doi:10.1063/1.4921984}
R.~C. Clay and M.~A. Morales,
\newblock J. Chem. Phys. {\bf 142}, 234103 (2015).

\bibitem{UmrTouFilSorHen-PRL-07}
C.~J. Umrigar, J.~Toulouse, C.~Filippi, S.~Sorella, and R.~G. Hennig,
\newblock Phys. Rev. Lett. {\bf 98}, 110201 (2007).

\bibitem{doi:10.1063/1.2908237}
J.~Toulouse and C.~J. Umrigar,
\newblock J. Chem. Phys. {\bf 128}, 174101 (2008).

\bibitem{messmer1969variational}
R.~P. Messmer,
\newblock Theor. Chim. Acta. {\bf 14}, 319 (1969).

\bibitem{messmer1970variational}
J.~H. Choi, C.~F. Lebeda, and R.~P. Messmer,
\newblock Chem. Phys. Lett. {\bf 5}, 503 (1970).

\bibitem{ye2017sigma}
H.-Z. Ye, M.~Welborn, N.~D. Ricke, and T.~Van~Voorhis,
\newblock J. Chem. Phys. {\bf 147}, 214104 (2017).

\bibitem{shea2018}
J.~A. Shea and E.~Neuscamman,
\newblock J. Chem. Phys. {\bf 149}, 081101 (2018).

\bibitem{blunt2017charge}
N.~S. Blunt and E.~Neuscamman,
\newblock J. Chem. Phys. {\bf 147}, 194101 (2017).

\bibitem{blunt2018charge}
N.~S. Blunt and E.~Neuscamman,
\newblock arXiv {\bf 1808.09549} (2018).

\bibitem{morales2012msj}
M.~A. Morales, J.~McMinis, B.~K. Clark, J.~Kim, and G.~E. Scuseria,
\newblock J. Chem. Theory Comput. {\bf 8}, 2181 (2012),
\newblock PMID: 26588949.

\bibitem{Griewank-Walther-book}
A.~Griewank and A.~Walther,
\newblock {\em Evaluating Derivatives, Principles and Techniques of Algorithmic
  Differentiation, Second Edition},
\newblock Society for Industrial and Applied Mathematics, Philadelphia, 2008.

\bibitem{sorella2010qmcforces}
S.~Sorella and L.~Capriotti,
\newblock J. Chem. Phys. {\bf 133}, 234111 (2010).

\bibitem{Neuscamman2013jagp}
E.~Neuscamman,
\newblock J. Chem. Phys. {\bf 139}, 194105 (2013).

\bibitem{Aspuru-Guzik:2017:ad}
T.~Tamayo-Mendoza, C.~Kreisbeck, R.~Lindh, and A.~Aspuru-Guzik,
\newblock ACS Cent. Sci. {\bf 4}, 559 (2018).

\bibitem{KimQMCPACK2018}
J.~Kim et~al.,
\newblock J. Phys. Condens. Matter {\bf 30}, 195901 (2018).

\bibitem{Trail:2008:heavy_tail}
J.~R. Trail,
\newblock Phys. Rev. E {\bf 77}, 016703 (2008).

\bibitem{Trail:2008:alt_sampling}
J.~R. Trail,
\newblock Phys. Rev. E {\bf 77}, 016704 (2008).

\bibitem{QuantumPackage1.1}
{A. Scemama et al},
\newblock DOI: 10.5281/zenodo.825872.

\bibitem{MOLPRO_paper}
H.-J. Werner, P.~J. Knowles, G.~Knizia, F.~R. Manby, and M.~{Sch\"{u}tz},
\newblock WIREs Comput. Mol. Sci. {\bf 2}, 242 (2012).

\bibitem{shao:2015:qchem}
Y.~Shao et~al.,
\newblock Mol. Phys. {\bf 113}, 184 (2015).

\bibitem{scemama_2015_30624}
A.~Scemama, E.~Giner, T.~Applencourt, G.~David, and M.~Caffarel,
\newblock Quantum package v0.6, 2015.

\bibitem{MazzoliniE2715}
M.~Mazzolini et~al.,
\newblock Proc. Natl. Acad. Sci. U.S.A. {\bf 112}, E2715 (2015).

\bibitem{doi:10.1021/cr970155y}
A.~Natansohn and P.~Rochon,
\newblock Chem. Rev. {\bf 102}, 4139 (2002),
\newblock PMID: 12428986.

\bibitem{DASILVA20111}
C.~M. da~Silva et~al.,
\newblock J. Adv. Res. {\bf 2}, 1  (2011).

\bibitem{dugave2006cis-trans}
C.~Dugave,
\newblock {\em Cis-trans isomerization in biochemistry},
\newblock Wiley-VCH, Weinheim, 2006.

\bibitem{doi:10.1063/1.475804}
I.~Frank, J.~Hutter, D.~Marx, and M.~Parrinello,
\newblock J. Chem. Phys. {\bf 108}, 4060 (1998).

\bibitem{doi:10.1080/00268970512331339378}
I.~T. *, U.~F. Röhrig, and U.~Rothlisberger,
\newblock Mol. Phys. {\bf 103}, 963 (2005).

\bibitem{doi:10.1002/pol.1965.110030616}
M.~Russo, L.~Mortillaro, C.~D. Checchi, G.~Valle, and M.~Mammi,
\newblock J. Polym. Sci. B {\bf 3}, 501 (1965).

\bibitem{doi:10.1146/annurev.pc.34.100183.000335}
D.~J. Clouthier and D.~A. Ramsay,
\newblock Annu. Rev. Phys. Chem. {\bf 34}, 31 (1983).

\bibitem{doi:10.1063/1.2741534}
M.~Burkatzki, C.~Filippi, and M.~Dolg,
\newblock J. Chem. Phys. {\bf 126}, 234105 (2007).

\bibitem{ROOS1995215}
B.~O. Roos and K.~Andersson,
\newblock Chem. Phys. Lett. {\bf 245}, 215  (1995).

\bibitem{GHIGO2004142}
G.~Ghigo, B.~O. Roos, and P.~Åke Malmqvist,
\newblock Chem. Phys. Lett. {\bf 396}, 142  (2004).

\bibitem{hoy1979precise}
A.~R. Hoy and P.~R. Bunker,
\newblock J. Mol. Spectrosc. {\bf 74}, 1 (1979).

\end{thebibliography}

\end{document}